\documentstyle[12pt,equations,epsf]{article}

\def\dcd{\Delta_c}
\def\dsd{\Delta_s}
\def\thm{\theta^-}
\def\thp{\theta^+}
\def\cp{c_+}
\def\sp{s_+}
\def\sm{s_-}
\def\cm{c_-}
\def\php{P_H^+}
\def\phm{P_H^-}
\def\pvp{P_V^+}
\def\pvm{P_V^-}
\def\ppt{P_T^+}
\def\pmt{P_T^-}
\def\plp{P_L^+}
\def\plm{P_L^-}

\def\lvert{\left|}
\def\rvert{\right|}

\def\what{\widehat}

\def\bm{b_\mu}

\def\cfsq{c_f^2}
\def\dfsq{d_f^2}

\def\cmusq{c_\mu^2}
\def\dmsq{d_\mu^2}

\def\afbsqpbfsq{\left(a_f^2 \beta_f^2+b_f^2\right)}
\def\amb{a_\mu}
\def\ambsqpbmsq{\left(a_\mu^2+b_\mu^2\right)}
\def\ambsqmbmsq{\left(a_\mu^2-b_\mu^2\right)}
\def\Pig{\Pi_\gamma}

\def\ImPiz{\Im(\Pi_Z)}

\def\ModPihsq{\lvert\Pi_h\rvert^2}
\def\ModPizsq{\lvert\Pi_Z\rvert^2}

\def\czeta{\cos\zeta}
\def\szeta{\sin\zeta}

\def\c3thcut{\cos\left(3\theta_{cut}\right)}

\newcommand{\nc}{\newcommand}

\nc{\fdiag}{0}
\nc{\bg}{B. Grzadkowski}
\nc{\BG}{Bohdan Grzadkowski}
\nc{\lsp}{\;\;\;\;\;\;\;\;}
\nc{\beq}{\begin{equation}}   \nc{\eeq}{\end{equation}}
\nc{\bea}{\begin{eqnarray}}   \nc{\eea}{\end{eqnarray}}
\nc{\baa}{\begin{array}}      \nc{\eaa}{\end{array}}
\nc{\bit}{\begin{itemize}}    \nc{\eit}{\end{itemize}}
\nc{\ben}{\begin{enumerate}}  \nc{\een}{\end{enumerate}}
\nc{\bce}{\begin{center}}     \nc{\ece}{\end{center}}
\nc{\non}{\nonumber}
\nc{\lumun}{\;{\hbox {fb}^{-1}}{\hbox {yr}^{-1}}}
\nc{\hc}{\hbox {h.c.}}
\nc{\re}{\hbox {Re}}
\nc{\im}{\hbox {Im}}
\nc{\etal}{\hbox{et al.}}
\nc{\pbarn}{\;\hbox {pb}}
\nc{\PRD}[3]{{\it Phys.\ Rev.}\ {{\bf D{#1}} (#2) #3}}
\nc{\PRL}[3]{{\it Phys.\ Rev.\ Lett.}\ {{\bf {#1}} (#2) #3}}
\nc{\plb}[3]{{\it Phys.\ Lett.}\ {{\bf B{#1}} (#2) #3}}
\nc{\npb}[3]{{\it Nucl.\ Phys.}\ {{\bf B{#1}} (#2) #3}}
\nc{\ptp}[3]{{\it Prog.\ Theor.\ Phys.}\ {{\bf {#1}} (#2) #3}}
\nc{\zfp}[3]{{\it Z.\ Phys.}\ {{\bf C{#1}} (#2) #3}}
\nc{\mpla}[3]{{\it Mod.\ Phys.\ Lett.}\ {{\bf A{#1}} (#2) #3}}
\nc{\RMP}[3]{{\it Rev.\ Mod.\ Phys.}\ {{\bf {#1}} (#2) #3}}
\nc{\ijmpa}[3]{{\it Int.\ J.\ of\ Mod.\ Phys.}\
               {{\bf A{#1}} (#2) #3}}
\nc{\app}[3]{{\it Acta\ Phys.\ Polon}\ {{\bf B{#1}} (#2) #3}}
\nc{\epj}[3]{{\it Eur. Phys. J.}\ {{\bf C{#1}} (#2) #3}}
\nc{\ra} {\rightarrow}
\nc{\cw}{\cos\theta_W}        \nc{\sw}{\sin\theta_W}
\nc{\ttbar}{t\bar{t}}
\nc{\bbbar}{b\bar{b}}
\nc{\tanb} {\tan \beta}
\nc{\twbdec} {t\rightarrow W^+ b}
\nc{\tbwbdec} {\bar{t} \rightarrow W^- \bar{b}}
\nc{\hprod} {e^+e^- \ra Z^\ast \ra H Z}

\def\tand{\tan\delta}
\def\ahat{\hat a}
\def\bhat{\hat b}
\def\ahatp{\hat a_0}
\def\bhatp{\hat b_0}
\def\call{{\cal L}}
\def\cals{{\cal S}}
\def\del{\delta}
\def\sigrts{\sigma_{\sqrt s}}
\nc{\epem} {e^+e^-}
\nc{\wpwm} {W^+W^-}
\nc{\tbar} {\bar{t}}
\nc{\bbar} {\bar{b}}
\nc{\wpp} {W^+}
\nc{\mt}{m_t}
\nc{\mts}{m_t^2}
\nc{\mw} {m_W}
\nc{\mws} {m_W^2}
\nc{\mz} {m_Z}
\nc{\mzs} {m_Z^2}
\nc{\mhs} {m_H^2}
\nc{\ma} {m_A}
\nc{\mas} {m_A^2}
\nc{\hdec}{H \ra t\bar{t}}
\nc{\ttbardec}{\ttbar \ra W^+W^-\bbbar}
\nc{\po}{\Phi_1}
\nc{\pod}{\Phi_1^\dagger}
\nc{\pht}{\Phi_2}
\nc{\phtd}{\Phi_2^\dagger}
\nc{\phtt}{{\tilde{\Phi}}_2}
\nc{\popo}{\po^\dagger\po}
\nc{\phtpt}{\pht^\dagger\pht}
\nc{\popt}{\po^\dagger\pht}
\nc{\phtpo}{\pht^\dagger\po}
\nc{\sq}{\sqrt{2}}
\nc{\nsd} {N_{SD}}
\nc{\ntt} {N_{tt}}
\nc{\vs}{\vspace{2mm}}
\nc{\sty}{\hat{S}^t_1} \nc{\pty}{\hat{P}^t_1}
\nc{\sts}{(\sty)^2}      \nc{\pts}{(\pty)^2}
\nc{\yts}{\sts+\pts}
\nc{\sby}{\hat{S}^b_1} \nc{\pby}{\hat{P}^b_1}
\nc{\sbs}{(\sby)^2}      \nc{\pbs}{(\pby)^2}
\nc{\ybs}{\sbs+\pbs}

\def\gam{\gamma}

\def\mupmum{\mu^+\mu^-}

\def\rts{\sqrt s}
\def\br{BR}

\def\h{h}
\def\mh{m_{\h}}
\def\gamh{\Gamma_{\h}^{\rm tot}}

\def\hh{H^0}
\def\ha{A^0}

\def\mhh{m_{\hh}}
\def\mha{m_{\ha}}

\def\lsim{\mathrel{\raise.3ex\hbox{$<$\kern-.75em\lower1ex\hbox{$\sim$}}}}
\def\gsim{\mathrel{\raise.3ex\hbox{$>$\kern-.75em\lower1ex\hbox{$\sim$}}}}
\def\anti{\overline}
\def\pbi{~{\rm pb}^{-1}}
\def\fbi{~{\rm fb}^{-1}}

\def\gev{\,{\rm GeV}}
\def\tev{\,{\rm TeV}}

\begin{document}
%
\font\fortssbx=cmssbx10 scaled \magstep2
\medskip
\begin{flushright}
$\vcenter{
\hbox{\bf UCD-99-23} 
\hbox{\bf IFT-99-31}
\hbox{\bf hep-ph/0003091}
\hbox{March, 2000}
}$
\end{flushright}
\vspace*{2cm}
\begin{center}
{\large{\bf How Valuable is Polarization at a Muon Collider? 
A Test Case: Determining the CP Nature of a Higgs Boson}}\\ 
\rm
\vspace*{1cm}
\renewcommand{\thefootnote}{\alph{footnote})}
{\bf \BG$^1$}
\footnote{E-mail:{\tt bohdang@fuw.edu.pl}} 
{\bf John F. Gunion$^2$}
\footnote{E-mail:{\tt jfgucd@physics.ucdavis.edu}} 
and {\bf Jacek Pliszka$^1$}
\footnote{E-mail:{\tt pliszka@fuw.edu.pl}}\\

\vspace*{1.5cm}
{$^1$ \it Institute of Theoretical Physics, Warsaw University, 
Warsaw, Poland}\\
{$^2$ \it Davis Institute for High Energy Physics, 
UC Davis, CA, USA }\\

\vspace*{1.5cm}

{\bf Abstract}
\end{center}
\vspace{5mm} 
We study the use of polarization asymmetries at a muon collider
to determine the CP-even and CP-odd couplings of a Higgs boson
to $\mupmum$. We determine achievable accuracy as
a function of beam polarization and luminosity. The appropriate
techniques for dealing with the polarization precession are
outlined. Strategies especially appropriate for
a two-Higgs-doublet model (including the MSSM) are given.
Our general conclusion is that polarization will be very useful,
especially if the proton source is such that full luminosity
in the storage ring can be retained even after imposing cuts
on the originally accepted muons necessary for $P\gsim 0.4$ 
for each beam.

\vfill
\setcounter{page}{0}
\thispagestyle{empty}
\newpage

\renewcommand{\thefootnote}{\sharp\arabic{footnote}}
\setcounter{footnote}{0}

\section{Introduction}

Although the origin of mass 
has still not been established, it is widely expected that electroweak
symmetry breaking is driven by elementary scalar dynamics, leading
to one or more physical Higgs bosons \cite{hhg}. It is also very possible
that CP violation arises either partially
or entirely as result of CP violation in the Higgs sector~\cite{weinberg}.
Even if not, CP violation in other sectors
of the theory can induce CP violation in the Higgs sector
at the loop level~\cite{cp-phases}. 
Thus, we anticipate that the direct determination
of the CP nature of each observed Higgs boson could be crucial to
unraveling the nature of the full theory. 

Even if the minimal
one-doublet Standard Model turns out to be nature's choice,
we will certainly want to know that the single observed Higgs boson
is entirely CP-even in nature. Alternatively, if
electroweak symmetry breaking turns out to be driven by technicolor-like
dynamics, it would be highly desirable
to be able to check that the narrow, light
pseudo-Nambu-Goldstone bosons (PGB's), that often
arise in such a theory and are rather Higgs-like
in many respects (see, for example, \cite{pgb}), 
do indeed have CP-odd coupling to fermions. 

The means for determining the CP nature of an observed narrow
resonance are limited. If it is determined that its $WW$
or $ZZ$ coupling is substantial (small), then we will know that it has
a substantial (small) CP-even component. But, 
even in the general two-Higgs-doublet model the relation between
this coupling and the CP-even and CP-odd fermionic couplings
(denoted $a$ and $ib\gamma_5$, respectively, in 
$\anti f(a+ib\gamma_5)f$\footnote{Phases can always be chosen so that,
for any given $f$, $a$
and $b$ are real.})
is model-dependent. However, polarization correlations can allow
one to extract the ratio $b/a$ in a model independent manner.
For a light resonance, one possibility is to employ
the $\tau^+\tau^-$ decay of the resonance. If the $t\anti t$ mode
is open, then it too can be employed. 
These final-state possibilities were examined for a muon collider
in \cite{gg,soni}. One can also look
for certain characteristic angular distributions in associated
production, $b\anti b+$resonance or $t\anti t+$resonance,
that are sensitive to $b/a$ \cite{gghgp}. 
However, the most elegant approach to determining
the CP nature of a neutral resonance
is to employ initial state polarization asymmetries.
This is possible only for
$\gam\gam$ \cite{gamgam}
or $\mupmum$ \cite{soni,bbgh} production of the resonance. But, the $\gam\gam$
coupling of a neutral resonance is the result of either loop(s) (in
the Higgs case) 
or an anomaly (in the PGB case), and is not a direct measure of 
the elementary fermionic couplings. This leaves $\mupmum$ collisions, which
are also the only way in which we will
probe 2nd generation fermionic couplings in models where the strength
of the fermionic coupling is proportional to the fermion mass
(implying that a resonance with mass $>2m_\tau$ will always decay
to 3rd generation fermions and any CP-odd coupling to two photons
 will be dominated by the top quark loop).

In the limit of $\beta_\mu=\sqrt{1-4m_\mu^2/m_R^2}\to 1$,
the cross section for production of a resonance, $R$, with 
$\anti\mu(a+ib\gamma_5)\mu$ coupling to the muon 
takes the form
\bea
\overline\sigma_S(\zeta)&=&\overline\sigma_S^0\left(
1+P_L^+P_L^-+P_T^+P_T^-\left[{a^2-b^2\over a^2+b^2}\cos\zeta-{2ab\over a^2+b^2}\sin\zeta \right]
\right)\nonumber\\
&=&\overline\sigma_S^0\left[1+P_L^+P_L^-+P_T^+P_T^-\cos(2\delta+\zeta)\right]\,,
\label{sigform}
\eea
where $\delta\equiv \tan^{-1}{b\over a}$ and
$P_T$ ($P_L$) is the degree of transverse (longitudinal)
polarization\footnote{Note that the 
cross section for longitudinally polarized muons
depends only on $a^2+b^2$ and cannot be used to extract information
about the CP nature of the resonance's couplings to muons.}
 of each of the beams defined as $P\equiv {f^+-f^-}$,
with $f^+$ being the fraction of muons with spin in the dominant
direction and $f^-$ the fraction of muons with opposite spin.
Here, $\zeta$ is the angle of the $\mu^+$ transverse polarization relative
to that of the $\mu^-$ as measured using the the direction of the $\mu^-$'s
momentum as the $\hat z$ axis. The $S$ subscript denotes `signal'.
Both the $\cos\zeta$ and $\sin\zeta$ dependences allow significant
sensitivity to the ratio of interest, $b/a$, even though
only the $\sin\zeta$ term is truly CP-violating. 
The value of $\overline\sigma_S^0$, which results from
convoluting the resonance shape with a Gaussian
distribution in $\rts$, depends upon the model, the detector, 
the final state mode and the machine parameters. Ignoring final
state efficiencies and acceptance cuts, and integrating
over final state phase space, one finds \cite{bbgh}
a result that can be approximated by\footnote{This form,
from Ref.~\cite{jfgtalks}, has 
the correct asymptotic limits for $\Gamma_R/\sigrts\to 0$ and
$\Gamma_R/\sigrts\to \infty$ and is always within $18\%$ of the
 exact result.}
\beq
\overline\sigma_S^0={4\pi\Gamma(R\to\mupmum)\br(R\to F)\over m_R^2
\Gamma_R^{\rm tot}\left[
1+{8\over\pi}\left(\sigrts\over\Gamma_R^{\rm tot}\right)^2\right]^{1/2}}\,,
\label{sig0form}
\eeq
where $\sigrts$ is the Gaussian resolution in $\sqrt s$.

It is expected that the muon collider will first be operated with
the relatively small natural polarization of $P\sim 0.2$ for the
$\mu^+$ and $\mu^-$ bunches, since this will allow maximal machine
luminosity and lead to the largest number of Higgs events. The
bunch polarizations
will be oriented in the plane of the final storage ring so that their 
precession as the muon bunches circulate 
(for roughly 1000 turns) can be used to
precisely determine the central energies of the muon bunches 
and their Gaussian energy spread \cite{pstareport,euroreport}.
(One measures the oscillations of the energies of the electrons
from the decaying muons, which oscillations depend very sensitively upon
both the muon energy and the Gaussian energy spread.)
From Eq.~(\ref{sigform}) and the fact that $|P_L^+P_L^-|$ and $|P_T^+P_T^-|$
are both $\leq 0.04$, it is clear that these observations will,
to an excellent approximation, provide a measurement of $\overline\sigma_S^0$.
(By choosing appropriate relative phases between the $\mu^+$ and
$\mu^-$ polarization angles, the polarization dependent terms
in Eq.~(\ref{sigform}) will average to zero --- see the next section.)
Thus, a very accurate measurement of $\overline\sigma_S^0$
in several channels $F$ can be performed and a reasonably accurate
measurement of $\Gamma_R^{\rm tot}$ by a three-point scan will be possible.
A model-independent determination of $a^2+b^2$ then becomes
possible from Eq.~(\ref{sig0form}) by computing  
$\Gamma(R\to \mupmum)$ using the measured values of $\overline\sigma_S^0$,
$\Gamma_R^{\rm tot}$, and $\sigrts$ and a determination of 
$\br(R\to F)$. To obtain $\br(R\to F)$  
requires using the missing mass technique in the $ZR$ final state
(in either $\epem$ or $\mupmum$ collisions) 
to measure $\sigma(ZR\to ZX)$ ($X=$ anything) and the measurement
of $\sigma(ZR\to ZF)$ to
compute $\br(R\to F)=\sigma(ZR\to ZF)/
\sigma(ZR\to ZX)$. Once the basic Higgs observations have been
performed so that the Higgs width, branching ratios and $a^2+b^2$
are well-determined, the next task will be to perform a model-independent
measurement of $b/a$.  
This require maximizing the influence of the polarization-dependent
terms in Eq.~(\ref{sigform}). Determining the best procedure for doing
so and estimating the accuracy with which this measurement can
be carried out is the main goal of this paper.

Any given final state $F$ will have a significant background. In general,
the cross section for the background is nearly independent of $\sigrts$.
If we integrate over final state configurations then the background
is independent of $P_T^+$ and $P_T^-$ and $\zeta$.
In order to roughly understand the level of sensitivity 
to $b/a$ that can be achieved,
let us for the moment imagine that we can set $P_L^+=P_L^-=0$
and choose $P_T^+=P_T^-=P_T$.
We then 
denote the integrated background cross section by $\overline\sigma_B^0$. 
We imagine isolating
${a^2-b^2\over a^2+b^2}$  and ${-2ab\over a^2+b^2}$, 
respectively, via the asymmetries
\bea
{\cal A}_I&\equiv& {\overline\sigma_S(\zeta=0)-\overline\sigma_S(\zeta=\pi) 
\over \overline\sigma_S(\zeta=0)+\overline\sigma_S(\zeta=\pi) }=
P_T^2{a^2-b^2\over a^2+b^2}=P_T^2\cos2\delta\,,\\
{\cal A}_{II}&\equiv& {\overline\sigma_S(\zeta=\pi/2)-\overline\sigma_S(\zeta=-\pi/2) 
\over \overline\sigma_S(\zeta=\pi/2)+\overline\sigma_S(\zeta=-\pi/2) }
=-P_T^2{2ab\over a^2+b^2}=-P_T^2\sin2\delta\,.
\label{asyms}
\eea
Assuming that $(a^2+b^2)$ must be measured at the same
time as the asymmetries, the error for the measurement
of either of the asymmetries
${\cal A}_I$ or ${\cal A}_{II}$ is given by 
\beq
[\delta{\cal A}]^2={
\anti\sigma_S^0+\anti\sigma_B^0+{\cal A}^2\left(\anti\sigma_B^0-
\anti\sigma_S^0\right)\over L\,[{\anti\sigma_S^0}]^2
}
\label{aerror}
\eeq 
where $L$ is the integrated
collision luminosity, assumed distributed equally between
the two $\zeta$ measurements and we have temporarily taken
$P_T$ to be the same for all collisions (which is not
actually the case, as will be discussed shortly).
If $(a^2+b^2)$ has already been very accurately determined using
initial resonance production measurements and 
the technique outlined below Eq.~(\ref{sig0form}), then the expression 
for $[\delta{\cal A}]^2$ takes the form
\beq
[\delta{\cal A}]^2={
\anti\sigma_S^0+\anti\sigma_B^0-{\cal A}^2
[\anti\sigma_S^0]^2(\anti\sigma_S^0+\anti\sigma_B^0)^{-1}
\over L\,[{\anti\sigma_S^0}]^2
}
\label{aerrorp}
\eeq 
In either case, if $P_T<0.5$, 
and since ${\cal A}^2\leq P_T^4$,
the ${\cal A}^2$ in the numerator of $[\delta{\cal A}]^2$ 
can be neglected, implying that ${\cal A}^2\over [\delta{\cal A}]^2$
is proportional to $P_T^4 L$. 

From this discussion, it
is apparent that the ideal situation would be to arrange for
the muons to be entirely transversely polarized at the interaction
point (IP) and to be able to adjust the angle between the
$\mu^-$ and $\mu^+$ polarizations to be fixed at one of the
four values $\zeta=0,\pi/2,\pi,3\pi/2$ for each collision.
However, this is not possible if we are interested in a very
narrow resonance, such as a light SM-like Higgs boson
which will have a width of just a few MeV. The reasons follow.
First, it is important to note that without
intervention any horizontal (i.e perpendicular to the magnetic
field of the storage ring) polarization
will precess about the magnetic field and
therefore rotate relative to the momentum direction. Consequently, the
amounts of transverse and longitudinal polarization 
at the time the bunch passes the IP will oscillate. This will
be described in more detail in the following section. The only
way to overcome this oscillation would be to have a section
of the storage ring devoted to compensating for this spin precession 
on a turn-by-turn basis.  However, this conflicts with the requirement
that we be able to measure the central energy of each muon bunch
circulating in the storage ring to 1 part in $10^6$ (and
the beam energy spread of the bunch to better than 1 part in $10^2$) 
using measurements
of the oscillations of the energy of the secondary electrons
from the muon decays as the bunches circulate and their spins
precess. A large number of turns in which the precession
is allowed to occur without compensation is required.
The above precisions are those needed
for a light Higgs resonance with a SM-like width (of a few MeV)
in order to be certain
of remaining rather precisely centered on the resonance peak
and knowing exactly how that resonance peak is being sampled.
For a resonance with width of a few hundred MeV or larger,
knowledge of the beam energies to 1 part in $10^4$ would be adequate.
Since magnetic fields would be known and be stable at this level,
direct measurement of the bunch energies would not be required
and one could consider turn-by-turn spin compensation
so as to achieve the ideal transverse configurations at the IP
for each collision. Here, we will assume that the storage ring
will not initially be built with the extra magnetic components
required for such compensation. 

Thus, regardless of whether the Higgs or other resonance
is broad or narrow, it is necessary to determine 
the effects of spin precession on the accuracy with which $\cos2\delta$
and $\sin 2\delta$ can be determined. 
In what follows, we develop a procedure whereby 
the luminosity needed to achieve a given accuracy is only about 50\% larger
when the spins are allowed to precess than if the spins
could be taken to be purely transverse for each collision.

\section{Polarization Precession at a Muon Collider}

\begin{figure}[t]
\leavevmode
\epsfxsize=5.3in
\centerline{\epsffile{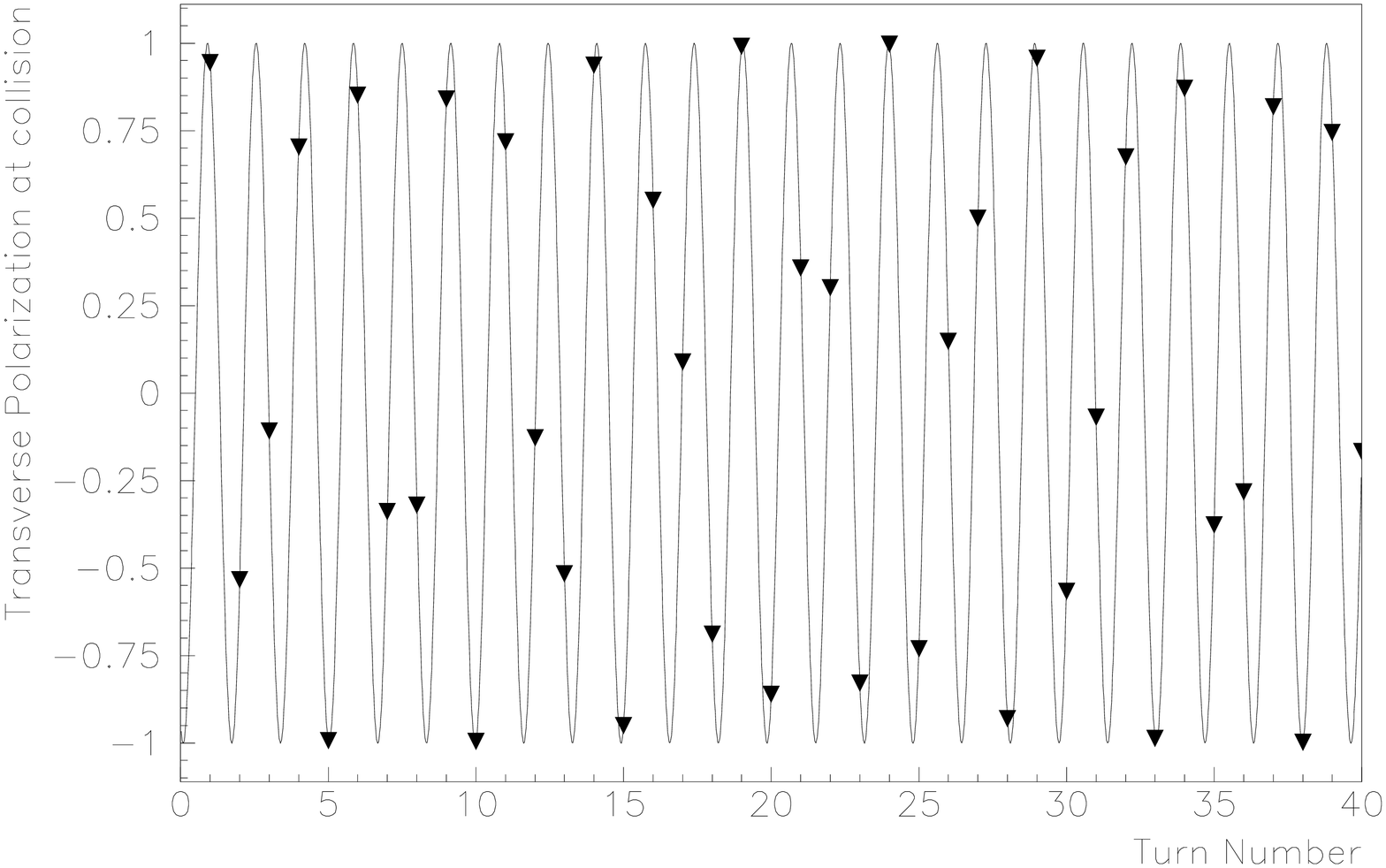}}
\medskip
\caption{We plot $P^{\perp}_H(N_T)/|\vec P_H|$ 
at the interaction point
as a function of the number of times, $N_T$, that the $\mu^-$ beam passes
the IP, assuming the muon collider is operating at a total center of 
mass energy of $\rts=110\gev$ and that the $\mu^-$
enters the storage ring with longitudinal polarization 
$P^{\parallel}_H/|\vec P_H|=1$. This plot is from \cite{rt}.}
\label{oscillations}
\end{figure}

Most muon collider designs are such that there are two bunches, each,
of $\mu^+$'s and $\mu^-$'s circulating in opposite directions in
the storage ring. Typically, these bunches will be stored for
about 1000 turns. 
Once the bunches enter the storage ring, any polarization
in the horizontal plane will precess in the vertical magnetic field
of the storage ring. Further, as noted in the introduction,
for the case of a very narrow resonance this precession
is needed to measure the bunch energies and Gaussian spreads to high
accuracy, implying that the polarizations of the bunches
should not be manipulated
once they are stored in the ring. The impact of such precession
on extracting physics has not been carefully examined to date.
Thus, we provide in this section 
a fairly detailed explanation of the considerations
and procedures that must be employed. Even though we shall focus
on the case of a narrow spin-0 resonance, the general features
of our discussion will have wider applicability.

The typical configuration will be such that
a given bunch enters the ring with a component $\vec P_V$ ($\vec P_H$)
of polarization vertical (horizontal) 
with respect to the plane of the storage ring. [$\vec P_V$ and $\vec P_H$
are defined in the muon rest frame; see Eqs.~(\ref{smum}) and (\ref{smup})
below.]
$\vec P_H^-$ and $\vec P_H^+$ will rotate in the same directions
as the $\mu^-$ and $\mu^+$ themselves (that is in opposite directions).
The rate of rotation of the $\vec P_H$'s as viewed from the laboratory frame
is somewhat different than the rate at which the bunches themselves
rotate. The mismatch
means that if, for instance, $\vec P_H^-$ were longitudinal at the
time the $\mu^-$ bunch first enters the storage ring, it will
not remain so but rather it will precess
into the transverse direction and then back to the opposite longitudinal
direction, and so forth. 

To be precise, we assume that the storage ring's magnetic field
points in the $-\what y$ direction: $\vec B=-B\what y$. The $\mu^-$
is then rotating clockwise about the $\what y$ axis. At any
given moment, we define $\what z=\what p_{\mu^-}$.
Then, $\what x$ points radially outward.
The angle $\theta^-$ is defined
as the angle by which one must rotate (in the $\mu^-$ rest frame)
about the $\what y$ axis in
order to get from $\what z$ to $\vec P_H^-$; $\theta^-$ is an `over-rotation'
angle in that it is the additional angle of rotation of the spin
as compared to the angle of rotation of $\vec p_{\mu^-}$.
After boosting to the laboratory frame, 
the complete four-component spin vector for the $\mu^-$ is then:
\beq
s_{\mu^-}=P_H^-\left[\gamma(\beta,\what z)\cos\theta^- - 
(0,\what x)\sin\theta^-\right]+
P_V^-(0,\what y)\,.
\label{smum}
\eeq
The standard result of Ref.~\cite{bargmanetal}, 
assuming that the $\mu^-$ beam enters
the storage ring with $\what P_H^-=\what p_{\mu^-}$, is
$\theta^-(N_T) =\omega (N_T-1/2)$,
where $N_T$ is the number of turns during storage,
counted starting with $N_T=1$ the first time the bunch
passes the IP ($N_T=1/2$ at bunch insertion), and
$\omega=2\pi \gamma {g_\mu-2\over 2}$, with $\gamma=E/m_\mu$ and
${g_\mu-2\over 2}=1.165924\times 10^{-3}$. One finds
that $\omega=\pi$ for a beam energy of $90.6223\gev/2$ (i.e.
close to $\mz/2$),
implying that an initially longitudinal $\mu^-$ horizontal 
polarization becomes transverse 
after the $\mu^-$ travels half way around the ring 
to the interaction point. However, for the
somewhat higher energies of interest for a Higgs factory, the degree
of horizontally transverse polarization (denoted by $\perp$) 
oscillates with $N_T$
according to $P^{\perp}_H(N_T)/|\vec P_H|=\sin \theta^{-}(N_T)$ as illustrated
in Fig.~\ref{oscillations}, where we have
chosen a convention in which $P_H^\perp$ is positive when it points
towards the center of the storage ring. We now discuss how
to take this oscillation into account. 

First, we note that similar results apply for the $\mu^+$. The $\mu^+$ will
be traveling in a counter-clockwise direction about
the $\what y$ axis; {\it at the interaction
point 1/2 way around the ring}, $\what p_{\mu^+}=-\what z$.
Once again, we can define an over-rotation angle $\theta^+$
(in the $\mu^+$ rest frame),
in terms of which the $\mu^+$'s spin vector (in the laboratory
frame) at the interaction point is written as
\beq
s_{\mu^+}=P_H^+\left[\gamma(\beta,-\what z)\cos\theta^+ - 
(0,\what x)\sin\theta^+\right]+
P_V^+(0,\what y)\,.
\label{smup}
\eeq
If we start
with $\what P_H^+=\what p_{\mu^+}$ at the time of insertion,
then $\theta^+(N_T) =\omega (N_T-1/2)$, just as for $\theta^-(N_T)$.
More generally, we can insert the $\mu^+$ beam with any initial
angle for $\what\php$ that we desire.

We assume that each time a Higgs event is observed we can compute,
or will have measured (prior to the IP, and
then extrapolated to the IP), the transverse polarizations of the bunches.
That is, we will know $\thm$ and $\thp$ for each interaction.
In fact, $\thp$ will be completely correlated with $\thm$, 
the correlation being determined
by the initial spin configuration with which
the bunches are injected into the storage ring.
We now give the expression for the cross section as a function of
$\thm$ and $\thp$, defining $\cm\equiv \cos \thm$ etc.,          
\bea
{\overline\sigma_S(\thp,\thm)\over\overline\sigma_S^0}&=&
(1+\php\phm\cp\cm)+\cos 2\delta(\pvp\pvm+\php\phm\sp\sm)
\nonumber\\
&\phantom{=}&+{\sin2\delta}(\phm\pvp\sm-\php\pvm\sp)\,.
\label{sigtheta}
\eea
Note that we obtain Eq.~(\ref{sigform}) using the obvious
replacements: $\php \cp=\plp$,
$\phm\cm=\plm$, $\pvp\pvm+\php\phm\sp\sm=\ppt\pmt\cos\zeta$, and
$\phm\sm\pvp-\php\sp\pvm=-\ppt\pmt\sin\zeta$. We also
note that if we choose $\pvp=\pvm=0$ and insert the bunches
so that $\thp=\thm+\pi/2$, then all polarization dependent terms
in Eq.~(\ref{sigtheta}) average to zero after many turns.
This is the configuration that would normally be employed 
for the initial Higgs resonance scans so that
knowledge of $b/a$ would not be needed 
in order to properly interpret these measurements.

However, to determine $b/a$ we will wish to employ one or more different
polarization configurations and
retain maximum information by binning Higgs events according to 
$(\thm,\thp(\thm))$. After integrating over  
$(\theta,\phi)$ configurations in an $f\anti f$ final state, one finds
(in the limit of $m_f/\rts\to 0$)
\bea
\sigma_B(\thm,\thp)&\propto& e_f^2e_\mu^2(1-P_H^+P_H^-\cp\cm)\Pi_\gam^2
\nonumber\\
&+&2e_fe_\mu c_f\left[c_\mu((1-P_H^+P_H^-\cp\cm)+d_\mu(P_H^-\cm-P_H^+\cp)
\right]\Pi_\gam\Re(\Pi_Z)
\nonumber\\
&+&(c_f^2+d_f^2)\left[(c_\mu^2+d_\mu^2)(1-P_H^+P_H^-\cp\cm)
+2c_\mu d_\mu(P_H^-\cm-P_H^+\cp)\right]|\Pi_Z|^2\,,\nonumber\\
\label{bform}
\eea 
where the $\gam,Z$ propagators and couplings are 
$\Pi_\gam$ and $\Pi_Z$ and $ie_f$ and $i\gam_\mu(c_f+d_f\gam_5)$,
respectively; see the Appendix for more details.

In order to give a simplified discussion, let us 
assume that $\overline\sigma_S^0$
(and, thence, $a^2+b^2$) has been precisely determined by first
resonance production  measurements
that do not focus on determining the Higgs CP properties.
As we have described in the introduction, this is very likely to be the case.
Then, remembering that $\thp$ can be considered to be a function of $\thm$
for any given choice of storage ring insertion configuration,
we can write the signal cross section as 
\beq
\overline\sigma_S^C(\thm)=f_0^C(\thm)+\cos2\delta\, f_c^C(\thm)+\sin2\delta\, 
f_s^C(\thm)\,,
\label{sigthm}
\eeq
where 
\bea
f_0^C(\thm)&=&\overline\sigma_S^0(1+\php\phm\cp\cm)\,,\nonumber\\
f_c^C(\thm)&=&\overline\sigma_S^0(\pvp\pvm+\php\phm\sp\sm)\,,\nonumber\\
f_s^C(\thm)&=&\overline\sigma_S^0(\phm\pvp\sm-\php\pvm\sp)\,
\label{fdefs}
\eea
depend upon the configuration, $C$, 
chosen for $\php,\phm,\pvp,\pvm,\thp(\thm)$. (Recall that $\thp(\thm)$
is determined by the choice made for the relative angle between the
polarization of the $\mu^+$ as compared
to that of the $\mu^-$ at the time of bunch insertion.)
The assumption that $a^2+b^2$ has been precisely determined already
corresponds to assuming that $f_0^C(\thm)$ is completely known.
Then, writing $\Sigma^C(\thm)\equiv\overline\sigma_S^C(\thm)+
\overline\sigma_B^C(\thm)$,
the $\Delta\chi^2$ difference between two different Higgs
models with the same $a^2+b^2$ but different values of $\delta$
will be given by
\beq
\Delta\chi^2=\sum_C 
L_C\left[\dcd^2 M_{cc}^C+2\dcd\dsd M_{cs}^C+\dsd^2 M_{ss}^C\right]
\,,
\label{dchisqgeneral}
\eeq
where  $L_C$ is the luminosity devoted to configuration $C$,
\beq
M_{cc}^C=\int {d\thm\over 2\pi} {[f^C_c(\thm)]^2\over\Sigma^C(\thm)}\,\quad
M_{cs}^C=\int {d\thm\over 2\pi} 
{[f^C_c(\thm)f^C_s(\thm)]\over \Sigma^C(\thm)}\,\quad
M_{ss}^C=\int {d\thm\over 2\pi} {[f^C_s(\thm)]^2\over \Sigma^C(\thm)}\,,
\label{mforms} 
\eeq
and we have defined $\dcd\equiv \Delta\cos2\delta$
and $\dsd\equiv \Delta\sin 2\delta$ (the differences in
these two quantities between the two models).\footnote{We note
that in the limit of just two bins corresponding to $\sin\zeta=\pm 1$
or $\cos\zeta=\pm 1$, $\Delta\chi^2=1$ corresponds to errors
for $\sin 2\delta$ or $\cos 2\delta$, respectively, as given in
Eq.~(\ref{aerrorp}).}
The $\thm$ integral is over the value of $\thm$ 
(and the correlated $\thp(\thm)$ value) at the interaction point.
Here, we are approximating the sum over the discrete $\thm$ values
that arise during the course of 1000 turns of the bunches
by an integral over $\thm$. This is an excellent approximation
unless $\rts/90.62\gev$ (where $90.62\gev$ is
the $\rts$ value such that $\thm$ is constant) 
is a ratio of small (compared to 1000) integers.

We will demonstrate below that the configurations can be chosen sufficiently
cleverly that one can neglect the $\thm$, and, indeed, the
entire configuration dependence of
$\Sigma^C(\thm)$ and approximate $\Sigma^C(\thm)\sim\overline\sigma_S^0+
\overline\sigma_B^0$, the sum of the unpolarized cross sections.
Under these circumstances, 
the $\Delta\chi^2$ for discriminating between two
different models with the same $a^2+b^2$ value (i.e. same total
unpolarized rate) will then be given by
\beq
\Delta\chi^2={[\overline\sigma_S^0]^2\over \overline\sigma_S^0
+\overline\sigma_B^0}\sum_C L_C \what\cals_C^2\,,
\label{dchisqs}
\eeq
where
\beq
\widehat \cals_C^2=\Biggl\langle\left[\dcd(\pvp\pvm+\php\phm\sp\sm)
\nonumber\
+\dsd (\phm\pvp\sm-\php\pvm\sp)\right]^2\Biggr\rangle_C
\label{sangles}
\eeq
is a measure of our `sensitivity' to $\cos2\del$ and $\sin2\del$.
The averaging is over roughly 1000 turns of the bunches
in the storage ring. To an excellent approximation,
this average depends only on the relative
values of $\thp$ and $\thm$ at the time of bunch insertion.
We will now give results for $\what\cals^2$ for (a) $\thm=\thp$
(b) $\thm=\thp+\pi$ (c) $\thm=\thp+\pi/2$. The result
for (d) $\thm=\thp+3\pi/2$ is the same as for (c). We find
\bea
\what\cals^2_{a}&=& 
\dcd^2\left[{3\over 8}(\php\phm)^2+\php\phm\pvp\pvm+(\pvp\pvm)^2\right]
+{1\over 2}\dsd^2\left[\php\pvm-\phm\pvp\right]^2\,,\nonumber\\
\what\cals^2_{b}&=&
\dcd^2\left[{3\over 8}(\php\phm)^2-\php\phm\pvp\pvm+(\pvp\pvm)^2\right]
+{1\over 2}\dsd^2\left[\php\pvm+\phm\pvp\right]^2\,,\nonumber\\
\what\cals^2_{c}&=&\dcd^2\left[{1\over 8}(\php\phm)^2+(\pvp\pvm)^2\right]
+{1\over 2}\dsd^2\left[(\php\pvm)^2+(\phm\pvp)^2\right]\,.
\label{calscases}
\eea
We note that $M_{cs}=0$ when the $\thm$ dependence of $\Sigma^C$ is
neglected.

As noted earlier in the previous section, 
the ideal situation would be to be able to choose 
four basic relative orientations
of the $\mu^-$ and $\mu^+$ transverse polarizations with respect
to each other: $\zeta=0,\pi/2,\pi,3\pi/2$. Constant
$\zeta=0$ or $\zeta=\pi$ could be accomplished by setting $\php=\phm=0$
and $\pvp=\pvm$ or $\pvp=-\pvm$, respectively.  However, 
as noted in the previous section, some
degree of horizontal precessing polarization is necessary if we are
to be able to measure the energies of the muon bunches with the
accuracy of 1 part in $10^6$ needed for a very narrow Higgs resonance.
It is estimated that $\php=\phm\sim 0.05-0.1$ \cite{rt} is required
for the energy measurement.
(We will adopt the optimistic choice of $0.05$
in the remainder of this paper.)  
The rest can be placed in the vertical directions.

The spin precessions make
it impossible to maintain $\zeta=\pi/2$ or $\zeta=3\pi/2$ 
as the bunches circulate. The simplest thing that one can do
is to inject, say, the $\mu^-$ bunches with purely horizontal polarization,
and maximize the vertical polarization for the $\mu^+$ bunches
subject to the requirement that $\php=0.05$. Sensitivity
to both the magnitude and sign of $\sin 2\delta$ 
arises from the variation of the $\sin2\delta\phm\sm\pvp$
term in Eq.~(\ref{sangles}), which at various extremes samples
$\zeta=\pi/2$ {\it and} $\zeta=3\pi/2$. (We again emphasize the importance of
either calculating, starting from the insertion configuration,
or measuring, from the decay spectrum, the precession angles
associated with each observed Higgs event.)

Very specifically, we thus consider the following configurations,
keeping in mind that the net polarization $P$ for the $\mu^+$
and $\mu^-$ bunches will be essentially the same.
\begin{description}
\item{I:}
To approximate the $\zeta=0$ configuration, we choose
$\php=\phm=P_H=0.05$, $\thm=\thp$, $\pvp=\pvm=\sqrt{P^2-P_H^2}$. 
The sensitivity is then given by 
\beq
\what\cals_I^2=\what\cals_a^2=\dcd^2 \left[3P_H^4/8+P_H^2(P^2-P_H^2)+
(P^2-P_H^2)^2\right]\,.
\label{sensa}
\eeq
\item{II:}
To approximate the $\zeta=\pi$ configuration, we choose
$\php=\phm=P_H=0.05$, $\thm=\thp+\pi$, $\pvm=-\pvp=-\sqrt{P^2-P_H^2}$. 
The sensitivity is then given again by Eq.~(\ref{sensa}): $\what\cals_{II}^2=
\what\cals_I^2$.
\end{description}
We now justify, for the case of $C=I,II$, 
the approximation of taking $\Sigma^C\sim \overline\sigma_S^0
+\overline\sigma_B^0$, employed in obtaining Eq.~(\ref{dchisqs})
from Eqs.~(\ref{dchisqgeneral}) and (\ref{mforms}).
We note that for both configuration I and configuration II the background, 
see Eq.~(\ref{bform}), will depend only very weakly on $(\thm,\thp)$
simply because $P_H^-$ and $P_H^+$ are both small. Similarly, the
$\php\phm\cp\cm$ term in $f_0^C(\thm)$ 
and the $\php\phm\sp\sm$ term in $f_c^C(\thm)$ will be very small.
Further, $f_s^C(\thm)$ can be approximately neglected because both
$\php$ and $\phm$ are small.
Finally, we will sum over configurations I and II (with equal
luminosity weighting) in order to determine $\cos2\del$. This has
the important consequence that (using $|P_V^{\pm}|\simeq P$, see above)
the leading $P^4$ term in Eq.~(\ref{sensa}) is obtained 
from Eqs.~(\ref{dchisqgeneral}) and (\ref{mforms}) via the structure
\bea
\sum_{C=I,II} M_{cc}^C&\sim& P^4[\overline\sigma_S^0]^2\left(
{1\over \overline\sigma_B^0+\overline\sigma_S^0(1+P^2\cos2\delta )}
+{1\over \overline\sigma_B^0+\overline\sigma_S^0(1-P^2\cos2\delta)}\right)
\nonumber\\
&=&P^4[\overline\sigma_S^0]^2\left(
{2(\overline\sigma_B^0+\overline\sigma_S^0)\over 
(\overline\sigma_B^0+\overline\sigma_S^0)^2-P^4\cos^22\delta
(\overline\sigma_S^0)^2}
\right)\,,
\label{cancellation}
\eea
so that even for $P$ as large as $0.5$ or so (as we shall later consider)
the $P^4$ correction term in the denominator 
can be neglected.\footnote{Note the
analogy to Eq.~(\ref{aerrorp}) that is apparent when
we recall that $|\delta A|^2$ corresponds to $\Delta\chi^2=1$
and that ${\cal A}^2$ in the present case is $P^4\cos^22\delta$.} Thus, the
approximation of taking $\Sigma^C\sim \overline\sigma_S^0+\overline\sigma_B^0$
is quite accurate in this case.
\begin{description}
\item{III:}
To emphasize the $\zeta=\pi/2$ and $\zeta=3\pi/2$ configurations
over many turns of the bunches, we choose
$\phm=P$ ($\pvm=0$), $\php=P_H=0.05$ and $\pvp=\sqrt{P^2-P_H^2}$.
In addition, we choose $\thm=\thp+\pi/2$ (or $\thm=\thp+3\pi/2$)
so as to minimize dependence on $\dcd$.
The sensitivity is then given by
\beq
\what\cals_{III}^2=\what\cals_c^2={1\over 8}\dcd^2 P^2P_H^2+{1\over 2} \dsd^2 P^2(P^2-P_H^2)\,.
\label{sensc}
\eeq 
Note that if we choose either $\thm=\thp$ or $\thm=\thp+\pi$,
then the (undesired) sensitivity to $\dcd$ would be increased.
For $P_H\leq 0.05$ and $P>0.2$ (as will be the case), 
the $\dcd^2$ term can be dropped in Eq.~(\ref{sensc}).
\end{description}
In order to justify neglecting the dependence of $\Sigma^C$ on
$\thm$ in case III, we note that the terms of concern
in $\sigma_B(\thm,\thp)$, 
Eq.~(\ref{bform}), are those proportional to 
$(P_H^-\cm-P_H^+\cp)\sim P\cm$. The term of concern in $\sigma_S(\thm,\thp)$
is that proportional to $f_s(\thm)\sim \overline\sigma_S^0 P^2\sm$.
We will return to this latter term in a moment. First, we note that
the $P\cm$ term in $\sigma_B(\thm,\thp)$ 
can be approximately eliminated
without affecting $\sigma_S(\thm,\thp)$  by binning
together into a single $\cm^0$ bin 
events with the same sign of $\sm$ but
with $\cm=\cm^0$ and $\cm=-\cm^0$. (Even after
1000 turns, this is only an approximation since
the number of turns for which $\cm\sim\cm^0$ can differ significantly
from the number of turns for which $\cm\sim -\cm^0$,
depending upon the Higgs boson mass and the number of $\cm$ bins employed.)
Concerning the $\sigma_S(\thm,\thp)$ term proportional to $P^2\sm$,
we note that in computing $M_{ss}$, Eq.~(\ref{mforms}), the 
leading term in the numerator
is even in $\sm$ while the $P^2\sm$ correction term in 
the $\sigma_S(\thm,\thp)$ contribution to $\Sigma^C$ is odd. This means
that we will get a cancellation analogous to that in 
Eq.~(\ref{cancellation}) so that these corrections to our approximation
will be of relative order $P^4$ and can be neglected.

It is useful to note how these results compare to the ideal where
we imagine that $\zeta$ can be held fixed. For $\zeta=0,\pi$, our
sensitivity $\what\cals^2$ would be $\dcd^2 P^4$ while for $\zeta=\pi/2,3\pi/2$
it would be $\dsd^2P^4$. For $P_H=0.05$ we suffer a loss of about
$0.469$ (0.939) for $\dsd$ ($\dcd$) for $P=0.2$ and
$0.495$ (0.990) for $P=0.5$. In the remainder of this paper,
we approximate these $P_H=0.05$ results by assuming no loss
in $\what\cals^2$ for $\dcd$ and a factor of $1/2$ loss in
$\what\cals^2$ for $\dsd$. This latter factor must be overcome  
by increased luminosity to obtain the same statistical accuracy
as in the ideal case. To equalize sensitivity to $\dsd$ and $\dcd$,
we will assume that we accumulate twice as much luminosity for
the $\dsd$ configurations as for the $\dcd$ configurations.
Thus, for total integrated luminosity $L$ we accumulate $L/6$
in configuration I, $L/6$ in configuration II, and $2L/3$
in configuration III. Of the latter $2L/3$, 
as the spins precess $L/3$ will be accumulated
in configurations with $\sin\zeta<0$ and $L/3$ with $\sin\zeta>0$.
The sensitivity achieved is then equivalent to probing the four fixed $\zeta$
configurations ($\zeta=0,\pi/2,\pi,3\pi/2$) with $L/6$ each. The results
of the following sections will sometimes be phrased in this latter language.

\section{Maximizing Sensitivity to CP Violation}

The number and density of muons in
each bunch are limited by space-charge and muon beam power considerations.
Existing designs have proton source intensity such 
that these limits are saturated  
when momentum cuts on the initially accepted muons
coming from the target are such that
the muon beams have polarization of $P\sim 0.2$ (or less).
To date, there has been no strong reason for
designing the proton source so as to have
full luminosity (fully saturated bunches) for larger $P$ values.
This is because the typical design for the storage ring is such that
the polarization is longitudinal at the interaction point
half way around the ring, motivated by the fact that
(for longitudinal $P$) the signal rate is proportional to $1+P^2$,
while $\gam^*,Z^*$ induced (e.g. $b\anti b$) backgrounds are
proportional to $1-P^2$. 
In this configuration, the improvement of $S/\sqrt{S+B}$ with increasing
$P$ (assuming the integrated luminosity can be kept fixed) is 
rather slow,  
rising by $\lsim 20\%$ from $P=0.2$ to $P=0.4$. Thus, although
higher (longitudinal) $P$
would have some advantage if the proton intensity were such that
the two muon bunches were both fully saturated at $P\gsim 0.4$ rather than
at $P\sim 0.2$, it is usually accepted that this
advantage is not sufficient to justify the expense associated
with building a more intense proton source.

However, if determining the CP nature of the resonance's couplings
to $\mupmum$ is the goal, the (approximate) proportionality of 
the  $\what\cals^2$ sensitivity measures (see previous section)
to $P^4$ implies that retaining
maximal luminosity at $P\gsim 0.4$ has much greater advantages.
The extra proton source intensity required to achieve this is
determined by the fraction, $f_{\rm surv}(P)$,
of muons that survive after imposing
the momentum cuts required to achieve a certain $P$ (prior
to inserting the muon bunches into the storage ring). 
The variation of $f_{\rm surv}$
as a function of $P$ has been estimated by many different groups (see, e.g.,
\cite{pstareport,euroreport}).
Here, we adopt the post-cooling result given in \cite{pstareport}
(Fig. 21)
which is approximately described by $f_{\rm surv}(P)= 1.45-2.51P$ 
for $0.2\leq P\leq 0.6$. The number of muons in each bunch, $N_b$,
is proportional to the proton source intensity, $\call_{\rm ps}$,
times $f_{\rm surv}$. Using the specified form, one finds,
for example, that $f_{\rm surv}(0.45):f_{\rm surv}(0.39):f_{\rm surv}(0.2)\sim
1/3:1/2:1$, implying that $\call_{\rm ps}$ at $P=0.45$ (0.39) 
must be 3 (2) times that for $P=0.2$ to maintain full bunches.

If the machine is constructed with $\call_{\rm ps}$ only large enough
to saturate the bunches at $P=0.2$, then one finds that
$f_{\rm surv}$ is such that $S/\sqrt{S+B}$ declines
if one increases $P$ above 0.2 (and arranges the polarization
to be longitudinal at the interaction point).  However, maximum 
$\what \cals$ is rather achieved by increasing
$P$ to the point ($P\sim 0.39$) 
where each muon bunch has one-half of the $P=0.2$ number, $N_b$, and then
merging the two bunches into one bunch (prior to insertion into
the storage ring). Bunch merging effectively doubles the storage ring
luminosity at this $P$ compared to what would result
without bunch merging. (Without bunch merging, 
one would have luminosity proportional to $2\times (N_b/2)^2$
whereas with bunch merging the luminosity would be
proportional to $1\times N_b^2$.) As a result,
the collider luminosity at $P=0.39$ is only
a factor of 2 lower than at $P=0.2$ and
$P^2 \sqrt L$ is roughly a factor of 2.7 larger. (At the same time,
 $S/\sqrt{S+B}$ has declined by about a factor of 1.4.) 
If CP studies are the goal, one will choose the $P=0.39$ option
with bunch merging. If $\call_{\rm ps}$ larger than
that required to saturate the muon bunches at $P=0.2$ is available,
bunch merging may or may not be desirable for CP studies. For
$\call_{\rm ps}$ such that bunch saturation is achieved for
$P<0.42$ ($\geq 0.42$), the sensitivity $\what\cals$ for
CP studies will (will not) benefit by increasing $P$
still further to the point where the bunches can be merged.
If the bunch-merging option is appropriate, the integrated collider
luminosity $L$ will be one-half the (no merging) value at $P=0.2$.
If bunch saturation is the better option, $L$ will be exactly the
same as the $P=0.2$ value.
In either case, increasing $P$ beyond the bunch saturation or bunch merging
point (whichever is more optimal) causes $\what\cals$ to decline.

\begin{figure}[p]
\leavevmode
\epsfxsize=6.5in
\centerline{\epsffile{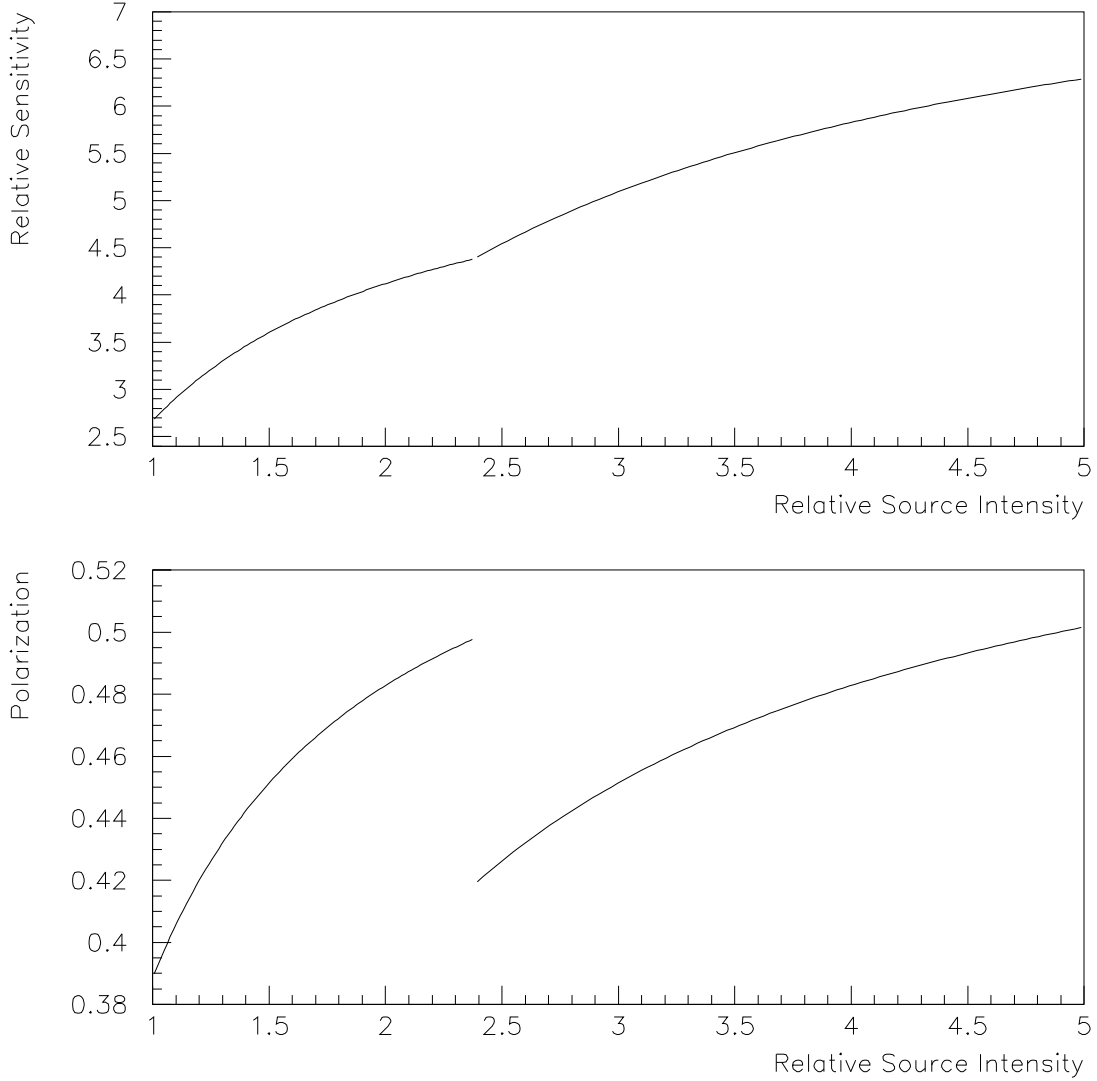}}
\medskip
\caption{We plot the maximum achievable 
relative sensitivity $R_{\cals}$ of the CP determination 
as a function of the relative proton source intensity $I$, both being relative
to values corresponding to two full $\mu^+$ and $\mu^-$ bunches (each)
at $P=0.2$. At $I=2.38$, one switches from bunch merging to bunch filling.
Also shown is the polarization being employed for a given $I$.} 
\label{sensitivity}
\end{figure}
 
To be more explicit, we define the relative sensitivity
$R_\cals\equiv\what\cals/\what\cals_0$, where the denominator
is that achieved using $P=0.2$ and $\call_{\rm ps}^0$ such that
the two bunches are saturated for $P=0.2$.  Let us also define
the relative source intensity $I\equiv \call_{\rm ps}/\call_{\rm ps}^0$.
In Fig.~\ref{sensitivity}, we plot $R_\cals$ obtained either by bunch merging
or bunch saturation (whichever gives the larger $R_\cals$) 
as a function of $I$.
Also shown is the corresponding polarization being employed as a function
of $I$. The corresponding functional forms are: for bunch merging,
$P^m(I)=a-{b\over 2I}$ with 
$R^m_\cals(I)=\left({P^m(I)\over 0.2}\right)^2 2^{-1/2}$;
for simply filling two bunches,
$P^f(I)=a-{b\over I}$ with $R^f_\cals(I)=\left({P^f(I)\over 0.2}\right)^2$.
Here, $a\simeq 0.577$ and $b\simeq 0.377$ for the form of $f_{\rm surv}$ given
earlier. Below (above) $I=2.38$ we employ $P^m$ and $R^m_\cals$
($P^f$ and $R^f_\cals$).
The results for $R_\cals$ show clearly that if determining the CP nature
of a resonance is the goal, then one gains by optimizing $P$
and having as large a proton source intensity as is feasible.

Of course, the absolute accuracy with which a measurement
of $b/a$ can be made is dependent upon the resonance model.
In the following section, we will consider several Higgs
boson examples and demonstrate that by using optimal polarization,
as described above, a very meaningful measurement of
$b/a$ can typically be performed.
Overall, we will conclude that the increase in our ability to determine
the CP nature of the muonic couplings of a resonance would provide
substantial motivation for spare proton source intensity.

Before ending this section, we note that there is a very 
interesting possibility
for increasing the number of muons with high polarization retained
after making the necessary selection cuts. In particular
\cite{pstareport,kaplan,geer}, 
if the accelerating gradient of the phase-rotation
device that immediately follows the pion capture solenoid is sufficiently
high ($\sim 4-5$ MV/m),  the correlation between the muon arrival time
and average muon polarization might be significantly enhanced,
resulting in a more effective selection of polarized muons, and a higher final
polarization for the same luminosity.
Such high-gradient phase-rotation designs are being actively considered
for the neutrino beam facility version of the muon accelerator and storage
ring~\cite{geer}.

\section{\boldmath Two test cases: a light Higgs with SM-like $b\anti b$
rate and a degenerate $\hh$--$\ha$ MSSM pair}

Consider first the case of a Higgs boson with SM-like $b\anti b$
final state rate and total width. For example, it might be that
a Higgs boson is detected and appears to have 
SM-like $WW/ZZ$ couplings and branching ratios to accessible final states,
in which case we will wish to determine if its fermionic couplings
are indeed CP-even.
Here, we assess the accuracy with which this verification can be
performed at a muon collider.  
As detailed in earlier studies \cite{bbgh,jfgtalks},
the result of Eq.~(\ref{sig0form}) 
and the very small width of a light SM-like 
Higgs boson means that the muon collider must be operated with the smallest
possible beam energy resolution, even if this results in
substantial luminosity sacrifice. For currently understood
designs, the best that can be achieved is
$\Delta E_{\rm beam}/E_{\rm beam}\equiv R$ 
with $R$ of order $R=3\times 10^{-5}$.\footnote{In this paper, all $R$
values will be quoted in absolute units and not in per cent.} 
For such an $R$ value, the yearly integrated
luminosity is anticipated to be of order $L=0.1\fbi$
when the bunches are full. We will examine
results achievable for $\mh=110\gev$ and $\mh=130\gev$
by employing only $F=b\anti b$.

The invariant amplitude squared for Higgs+$\gam^*+Z^*$ exchange is
given in the Appendix as a function of the relative angle $\zeta$
between the transverse polarizations of the $\mu^-$ and $\mu^+$
and as a function of both $\theta$ and $\phi$,
the angles describing the orientation of the $b$ and $\anti b$
momenta in the final state. We first note
that the interference term is never of importance for the cases
we explore here. Secondly, we observe that $\lvert{\cal M}\rvert^2$
from Higgs exchange is independent of $\theta$ and $\phi$, whereas that
for the $\gam^*+Z^*$ background does depend upon both $\theta$ and $\phi$.
Our analysis proceeds as follows.

We assume input values for
$(\hat a,\hat b)={(a,b)/\left(gm_\mu/2m_W\right)}$ 
($a\equiv a_\mu,b\equiv b_\mu$),
denoting them by $\ahatp$ and $\bhatp$. We integrate over $\phi$, but bin
events in $\cos\theta$ (bin label $j$).
As described, the spin precession means that we must also bin
events in $\thm$ (remembering that $\thp$
will always have a known correlation with $\thm$ for
a given spin-precession configuration); $\thm$
bins are labelled by $i$.
We devote $L/6$, $L/6$ and $2L/3$ to the spin-precession
configurations $C=I, II, III$ described in the previous section.\footnote{In
all cases studied, if there are no a priori
restrictions on the possible couplings, sensitivity to
$\hat b/\hat a$ is maximized by employing all the configurations.} 
The number of events
(including both signal and background) for a given 
configuration choice $C$, given $\thm$ bin $i$
and given $\cos\theta$ bin $j$ is denoted by $N(C,i,j)$.
The statistical error for each such bin is $\Delta N(C,i,j)=\sqrt{N(C,i,j)}$.
We then consider a different model characterized by $\ahat$
and $\bhat$ and compute
\beq
\Delta\chi^2=\sum_{C,\,i,\,j} 
{(N(C,i,j,\ahat,\bhat)-N(C,i,j,\ahatp,\bhatp))^2 
\over \Delta N^2(C,i,j,\ahatp,\bhatp)}\,.
\label{chisq}
\eeq
The discussion of Section 2 shows that with appropriate binning
and appropriate choices for our configurations we can effectively
drop the $C,i,j$ dependence of $\Delta N(C,i,j)$. The following
discussion will implicitly rely on this fact. 

\begin{figure}[p]
\leavevmode
\epsfxsize=5.5in
\centerline{\epsffile{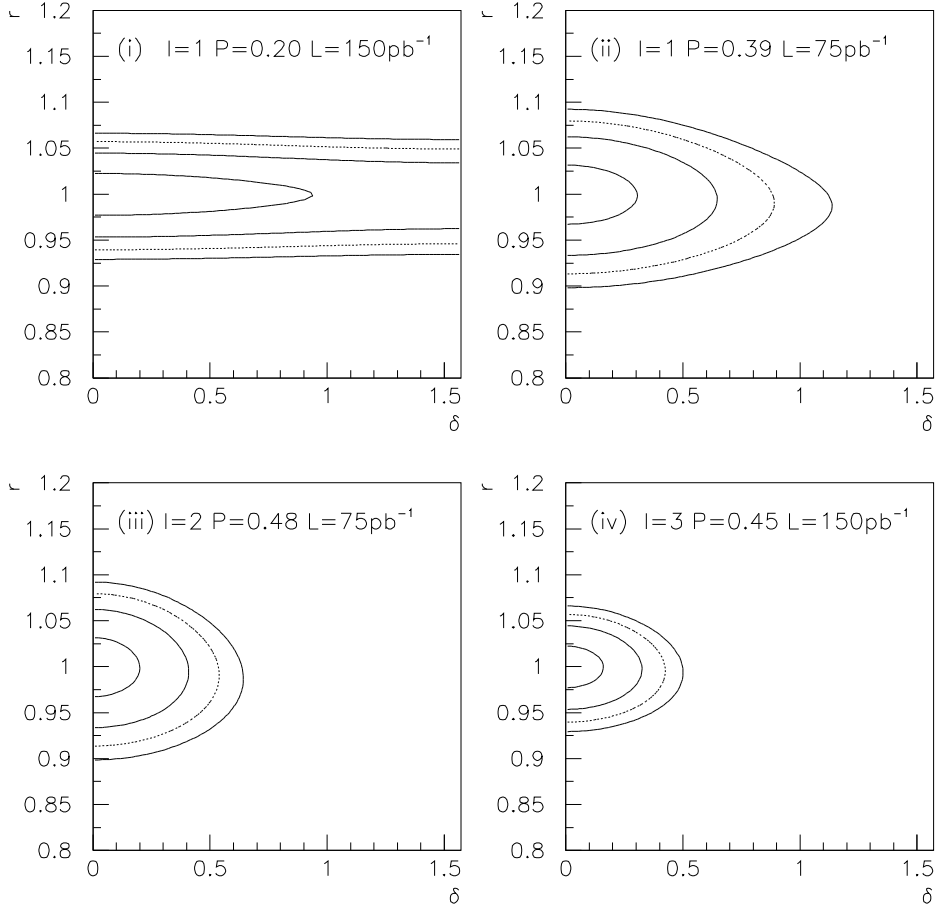}}
\medskip
\caption{We give contours at $\Delta\chi^2=1,4,6.635,9$ 
in the $\delta,r$ parameter space assuming that the integrated
luminosity at polarization $P=0.2$ (with full muon bunches) is $L=0.15\fbi$.
Four cases are compared: (i) $P=0.2$, $L=0.15\fbi$ (which
corresponds to $I=1$); (ii) maintaining same proton source intensity, $I=1$, 
but merging the bunches, corresponding $P^m(I=1)\sim 0.39$, for which
$L=0.075\fbi$; 
(iii) increasing the proton source intensity by a factor of two, $I=2$,
while merging the bunches, corresponding to $P^m(I=2)\sim 0.48$, for
which $L=0.075\fbi$; (iv) $I=3$, using just-full bunches, corresponding
to $P^f(I=3)\sim 0.45$, for which $L=0.15\fbi$. We assume a SM Higgs
boson ($\ahat_0=1$, $\bhat_0=0$)
with $\mh=110\gev$ and an overall efficiency (including $b$-tagging
efficiency) of $0.54$.}
\label{samplecasei}
\end{figure}

The accuracy with which $\ahat$ and $\bhat$ can be determined can
be assessed by drawing contours of constant
$\Delta\chi^2$ in $\delta\equiv \tan^{-1}\hat b/\hat a, 
r\equiv\sqrt{\hat a^2+\hat b^2}$ parameter space. 
The $n\sigma$ error in $\delta$ for example (which implicitly assumes
that $r$ is always
adjusted for a given $\hat b/\hat a$ so as to minimize
the $\Delta\chi^2$) is given by drawing the vertical line
tangent to the $\Delta\chi^2=n^2$ contour. We will also include
the $\Delta\chi^2=6.635$ contour corresponding to 99\% CL. 
The corresponding statistical errors 
for $r\equiv\sqrt{\hat a^2+\hat b^2}$
are obtained by the horizontal lines tangent to the contours.
However, the true errors for $r$
must incorporate the error coming from 
our imprecise knowledge of $\br(\h\to b\anti b)$ and $\gamh$.

Let us begin with the example of a SM Higgs boson ($\ahatp=1$, $\bhatp=0$)
with mass $\mh=110\gev$. To illustrate the importance of polarization,
we use as a reference point a total integrated luminosity 
at $P=0.2$ (with full muon bunches) of $L=0.15\fbi$ (per year). We plot
contours in $(\delta,r)$ parameter space for some sample cases:
(i) $P=0.2$, $I=1$ (i.e. nominal proton source intensity, $\call_{\rm ps}^0$);
(ii) $P=0.39$, $I=1$ (bunch merging at $\call_{\rm ps}^0$);
(iii) $P=0.48$, $I=2$ (bunch merging at $\call_{\rm ps}=2\call_{\rm ps}^0$);
(iv) $P=0.45$, $I=3$ (full bunches at $\call_{\rm ps}=3\call_{\rm ps}^0$).
The contours are presented in Fig.~\ref{samplecasei}. It is useful
to keep in mind
that $\bhat/\ahat=\tand$ is 1 for $\delta=\pi/4\sim 0.785$.
For the cases (i)-(iv) above, the
$1\sigma$, $2\sigma$, $99\%$ CL, and $3\sigma$ limits on $\delta$
are: (i) 0.94,$\ldots$; (ii) 0.30,0.64,0.89,1.14; (iii) 0.20,0.41,0.53,0.64;
and (iv) 0.15,0.32,0.42,0.50. The corresponding $1\sigma$ and $99\%$ CL
limits on $\hat b/\hat a=\tan\delta$ are:
(i) 1.36,$\ldots$; (ii) 0.31,1.23; (iii) 0.20,0.58; (iv) 0.15,0.45.
We see that even $I=1$ with bunch merging gives a reasonable $1\sigma$
measurement; however, a good 99\% CL limit on $\bhat/\ahat$ requires
$I\geq 2$. 

\begin{figure}[p]
\leavevmode
\epsfxsize=5.5in
\centerline{\epsffile{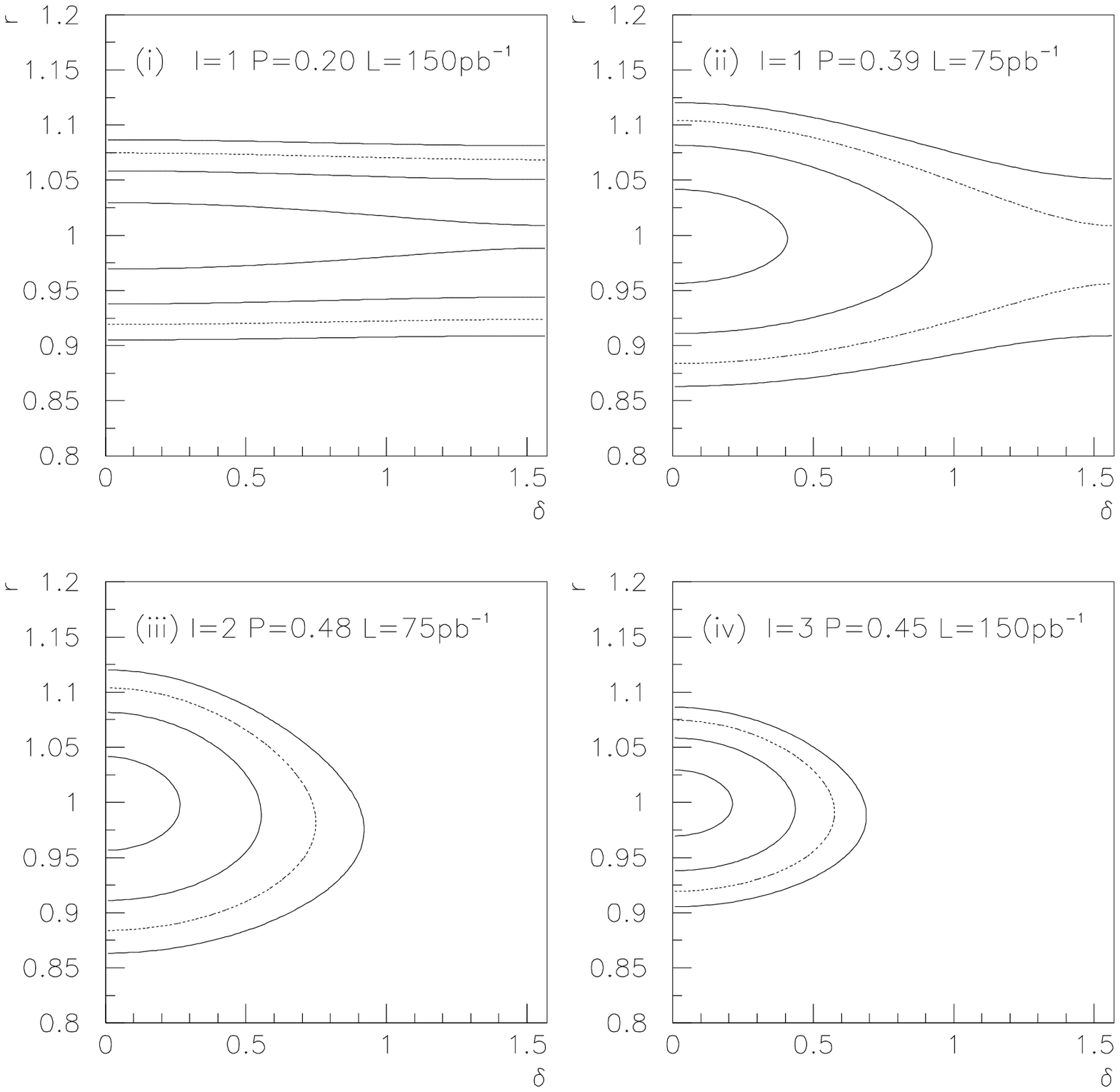}}
\medskip
\caption{Same as Fig.~\ref{samplecasei}, but for $\mh=130\gev$.}
\label{samplecaseii}
\end{figure}

\begin{figure}[p]
\leavevmode
\epsfxsize=5.5in
\centerline{\epsffile{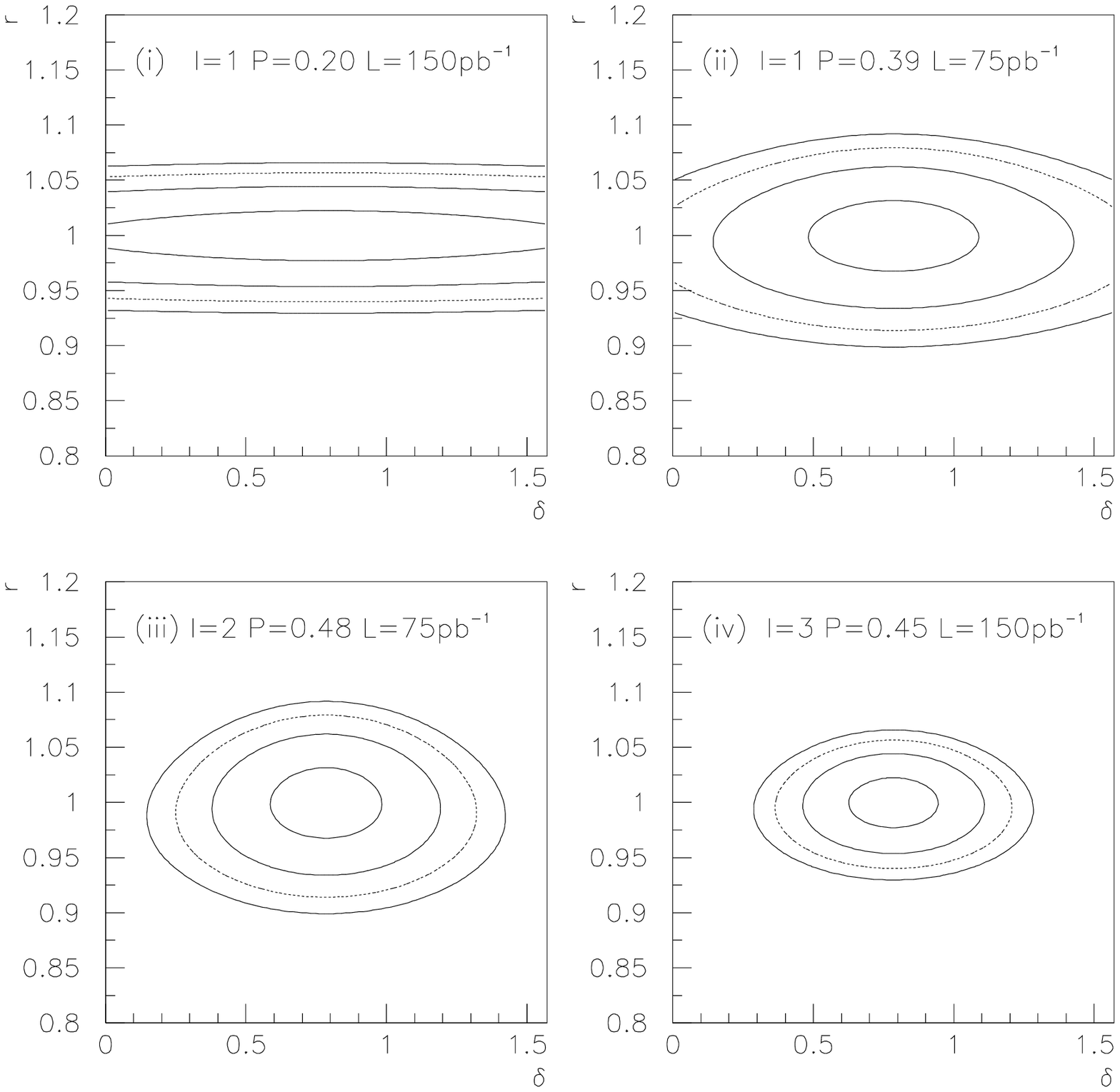}}
\medskip
\caption{Same as Fig.~\ref{samplecasei}, but
for a mixed-CP higgs boson with $\ahatp=\bhatp=1/\sqrt{2}$ and
$\mh=110\gev$.}
\label{samplecaseiii}
\end{figure}

Of course, as the Higgs mass increases, the event rates decline
and the results worsen.  This is illustrated in Fig.~\ref{samplecaseii}
which gives the same plots as Fig.~\ref{samplecasei}, but for $\mh=130\gev$.
Above $\mh=130\gev$, the $b\anti b$ branching ratio begins to decline
sharply and one should see a significant $WW^*$ final state rate,
which in itself would be a signal of a significant CP-even component
for the Higgs boson. A high discrimination 
check of the CP nature of the muonic
coupling would be possible by including the $WW^*$ channel
in the transverse polarization analysis. We will not pursue this here.

Another perspective on our errors is to ask how well separated
the $\bhat/\ahat$ measurement for a SM Higgs boson is from the
same measurement for an alternative Higgs boson of the general 
two-Higgs-doublet model (2HDM).  To illustrate, consider $\mh=110\gev$
with $\ahatp=\bhatp={1\over\sqrt{2}}$. The $\Delta\chi^2$ contours
for this input model are given in Fig.~\ref{samplecaseiii}. The $1\sigma$
contours are already non-overlapping
with those for a pure CP-even Higgs (Fig.~\ref{samplecasei}) 
for $P=0.39$ and $I=1$.  However,
the 99\% CL contours have a small overlap even at $I=3$ ($P=0.45$).
This again emphasizes the potential importance of larger-than-nominal
source luminosity for such discrimination.

\renewcommand{\arraystretch}{1.3}
\setlength{\tabcolsep}{0.03in}
\begin{table}[htb]
\centering
\caption{Event number pattern for different Higgs models as
a function of $\zeta$, assuming $P_L^\pm=0$ and $P_T^\pm=P$; see
Eq.~(\ref{sigform}).}
\vskip6pt
\begin{tabular}{|c|c|c|c|c|}
\hline
 $(\ahat,\bhat)$ & $\zeta=0$ & 
$\zeta=\pi/2$  & $\zeta=\pi$ & $\zeta=3\pi/2$ \\
[2pt]
\hline
$(1,0)$ & $1+P^2$ & 1 & $1-P^2$ & 1 \\
$(1/\sqrt2,1/\sqrt2)$ & 1 & $1-P^2$ & 1 & $1+P^2$ \\
$(0,1)$ & $1-P^2$ & 1 & $1+P^2$ & 1 \\
$(1/\sqrt 2,0)+(0,1/\sqrt 2)$ & 1 & 1 & 1 & 1 \\
\hline
\end{tabular}
\label{patterns}
\end{table}

Let us now turn to a situation in which there 
are two degenerate Higgs bosons. This can arise in the 2HDM
and also in the MSSM.
The rates as a function of the relative angle between the transverse
polarizations, $\zeta$, will depend in detail upon the
$\ahat$ and $\bhat$ values of the two Higgs bosons. As an example,
suppose that we have  a degenerate pair at
$\mh=110\gev$, one of which is pure CP-even with $\ahat=1/\sqrt2,\bhat=0$
($\delta=0$) and the other pure CP-odd with $\ahat=0,\bhat=1/\sqrt2$
($\delta=\pi/2$). Here, we have chosen the normalizations so that
the total unpolarized (i.e. averaged over the four $\zeta$ settings) 
production rate is the same as for a SM Higgs boson,
with each of the two Higgs contributing equally to this rate. (We will
also assume that the two Higgs bosons have the same $b\anti b$ branching
ratio as a SM Higgs boson.) In this situation, Eq.~(\ref{sigform})
shows that the production rate
will have no dependence on $\zeta$ [$\cos(0+\zeta)+\cos(\pi+\zeta)=0$], 
whereas any single Higgs model
will exhibit a distinct pattern as a function of $\zeta$, as illustrated
in Table~\ref{patterns}.   
Were we able to accumulate events at fixed $\zeta$, 
the pattern of $\Delta\chi^2$
for discriminating any two models from one another (assuming
all have the same $b\anti b$ production rate) is apparent
in the approximation where the $\zeta$ dependence of
the $\Delta N(\zeta)$ errors is neglected (as
approximately appropriate given the dominance of the
background contribution to $\Delta N$ and the weak
dependence of the background on $\zeta$); 
$\Delta\chi^2$ would simply be proportional to the
squares of the rate differences between two models 
summed over the four $\zeta$ values.
In the case where we compare the $\zeta$--independent rate
predicted in a degenerate Higgs pair
model to expectations for any given single Higgs boson, it is apparent
from the table that the $\Delta\chi^2$ between
the degenerate model and {\it any} single Higgs model is
independent of the latter. 
(This is true for arbitrary $\delta$ for a single Higgs
since $\cos^2(2\delta)+\cos^2(2\delta+\pi/2)+\cos^2(2\delta+\pi)+
\cos^2(2\delta+3\pi/2)=2$.)

However, because of spin precession, we must actually employ 
Eq.~(\ref{sigtheta}). 
Since $\sin 2\delta=0$ for both Higgs bosons while $\cos 2\delta=1$
for the CP-even and $\cos 2\delta=-1$ for the CP-odd Higgs boson,
after summing over both, 
the cross section will be the same as that obtained
if we set $\cos2\delta=0$ as well as $\sin 2\delta=0$.
To compare to the simplified discussion of the previous paragraph, 
let us recall that for our choices of configurations
and binning we can approximately
neglect the $C,i,j$ (i.e configuration, $\thm$ and $\theta$) dependence of 
the $\Delta N(C,i,j)$ denominators in Eq.~(\ref{chisq}). Then,
for the $L/6$, $L/6$, $2L/3$ luminosity weightings
for configurations $C=I$, $C=II$, $C=III$, respectively, the effective
sensitivity for discriminating between the 
[$\cos2\delta=\sin2\delta=0$]---equivalent
situation for the degenerate pair and the results expected for
any single Higgs boson characterized by angle $\delta$ 
will be proportional to ${1\over 3}\what\cals_a^2+{2\over 3}\what\cals_c^2
\sim {1\over 3} P^4\left(\dcd^2+\dsd^2\right)=
{1\over 3}P^4 \left(\cos^2 2\delta+\sin^2 2\delta\right)={1\over 3}P^4$ 
[see Eqs.~(\ref{sensa}) and (\ref{sensc})], 
i.e. again independent of the $\delta$ value
for any single Higgs boson to which one might compare.
One finds that
this approximation is actually quite good.
Taking $L=0.15\fbi$ for $I=1$ and $P=0.2$ and assuming a SM-like
$b\anti b$ production rate, we find $\Delta\chi^2=$ (i)
0.38; (ii) 2.8; (iii) 6.4; (iv) 9.8, for the four $(I,P)$
scenarios defined earlier, essentially independent
of the type of single Higgs boson exchange to which one compares.
Note that we get $\geq 99\%$ CL
exclusion of a single Higgs model only for $I>2$
if nature chooses a degenerate pair. In any case, the $\zeta$ dependence
is key to separating a degenerate CP-even plus CP-odd pair of Higgs
bosons from a single Higgs boson of any given CP nature.

\begin{figure}[p]
\leavevmode
\epsfxsize=5.5in
\centerline{\epsffile{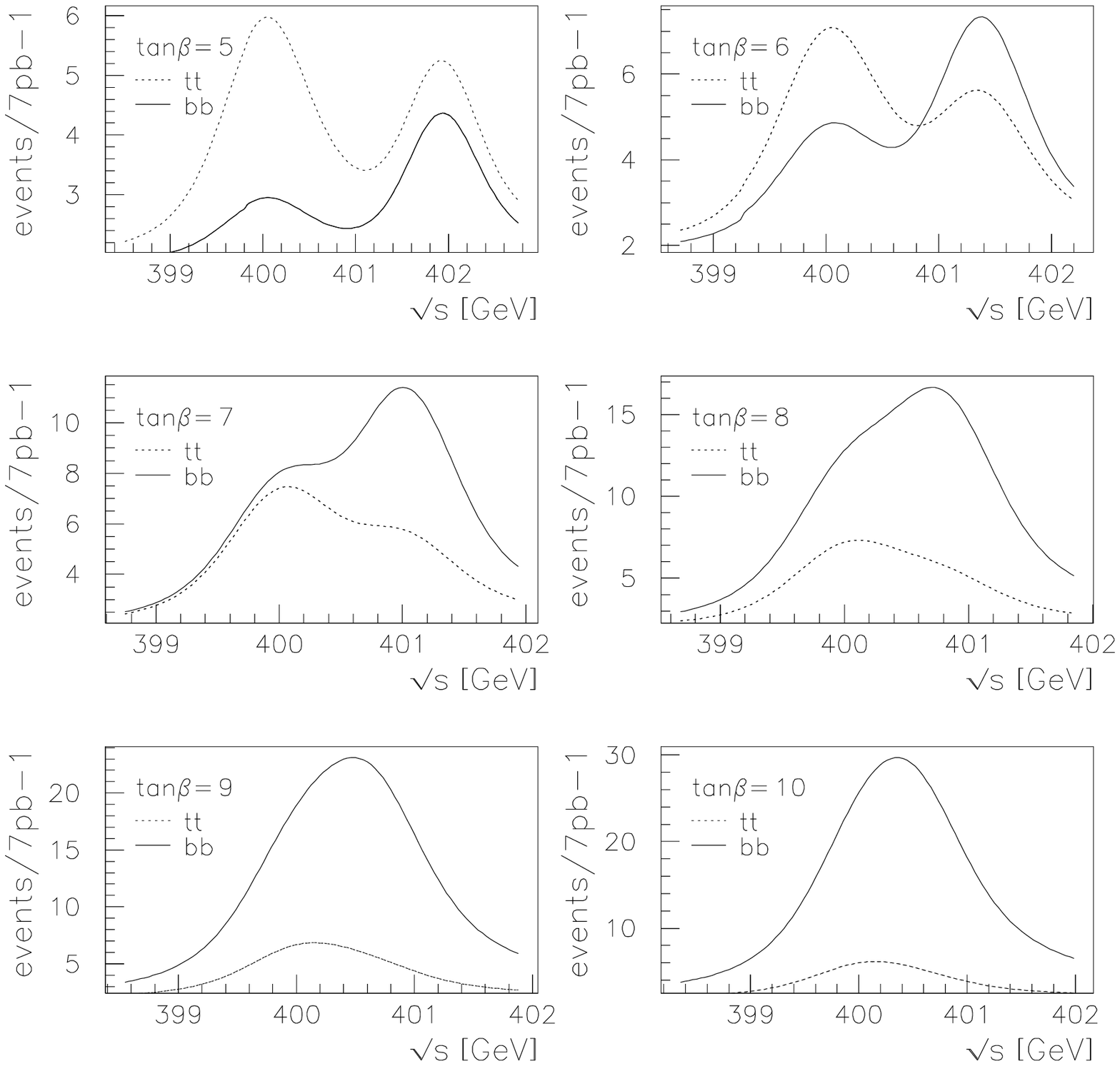}}
\medskip
\caption{We plot $b\anti b$ (solid) and $t\anti t$ (dashed) event rates
for total integrated luminosity of $L=7\pbi$ coming from
$\mupmum\to\hh+\ha$ as a function of $\rts$, assuming $\mha=400\gev$.
Each window is for the specific $\tanb$ value noted. These event
rates are to be multiplied by a factor of 1000 for the expected
yearly integrated luminosity of $7\fbi$. We employ squark masses
of $1\tev$ and no squark mixing. Supersymmetric decay channels are
assumed to be closed.}  
\label{degeneracy}
\end{figure}

In the MSSM, in the absence of CP violation the most likely situation
in which we will encounter a highly degenerate pair of Higgs bosons 
that cannot be easily separated by scanning in $\sqrt s$ is
in the limit of large $\mha$ and large $\tanb$.
The increasing degeneracy with increasing $\tanb$ is illustrated
for $\mha=400\gev$ in Fig.~\ref{degeneracy},
assuming squark masses of $1\tev$ and no squark mixing and a 
beam energy spread of $R=0.001$. 
Since the total widths of the $\hh$ and $\ha$
are substantial ($>1\gev$) for the $\mha$
and $\tanb$ values being considered, it is not
guaranteed that we will be able to separate the peaks.
The figure shows that we are able to observe two separate peaks (the $\ha$
peak being at lower mass than the $\hh$ peak) for moderate
$\tanb\lsim 6$.  But, for higher $\tanb$ values
the peaks begin to merge; for $\tanb\gsim 8$, $|\mhh-\mha|<1\gev$ and one
sees only a single merged peak.  The picture changes if squark mixing
is substantial; for instance, for $\mha=300\gev$, squark
masses of $1\tev$ and large squark mixing ($A_t=A_b=3\tev$), 
the $\hh$ and $\ha$ peaks actually cross at $\tanb\sim 5$.

To explore the various possible scenarios, we
begin by considering our ability to discriminate between 
an exactly degenerate pair of CP-even and CP-odd Higgs
bosons vs. a single Higgs boson (of a given type) as a function of $\tanb$.
We note that the present situation 
is significantly more favorable than that discussed above
with a light degenerate pair with SM-like widths, branching
ratios and production rate. Since the widths
of the $\hh$ and $\ha$ are significant, one
can operate the muon collider with 
the natural beam energy resolution of order $R=0.001$ and still
have $\sigrts<\Gamma$, which maximizes the
Higgs production rate.\footnote{Of course, one could possibly separate
the two Higgs bosons by employing $R<0.0001$, but it is unlikely
that the machine would be operated in this way at this higher energy
unless a Higgs resonance peak is seen and there is already
some evidence, through the techniques considered here, that 
there are actually two overlapping resonances with different
CP properties. Even then, there will remain the possibility 
of very close or even exact degeneracy.} At such $R$ and for $\sqrt s\sim
300\gev$ ($400\gev$), 
the nominal yearly integrated luminosity for full bunches at $P=0.2$
is estimated at $L\sim 2\fbi$ ($7\fbi$), i.e. more than
a factor of ten larger than the $R=3\times 10^{-5}$ value. 
Second, the $b\anti b$ branching
ratio for large $\mha$ and large $\tanb$ is inevitably of order $88\%$,
the only significant competitor being $\tau^+\tau^-$. 
Even for moderate $\tanb$ and for Higgs masses above $t\anti t$ threshold, 
the $b\anti b$ branching ratio 
remains substantial. Even more
importantly, at large $\mha$ the $\mupmum$ couplings of the $\hh$ 
and $\ha$ are enhanced relative to SM strength by a factor of $\tanb$
so that the $\mupmum$ branching ratio asymptotes to a constant.

\begin{figure}[p]
\leavevmode
\epsfxsize=5.5in
\centerline{\epsffile{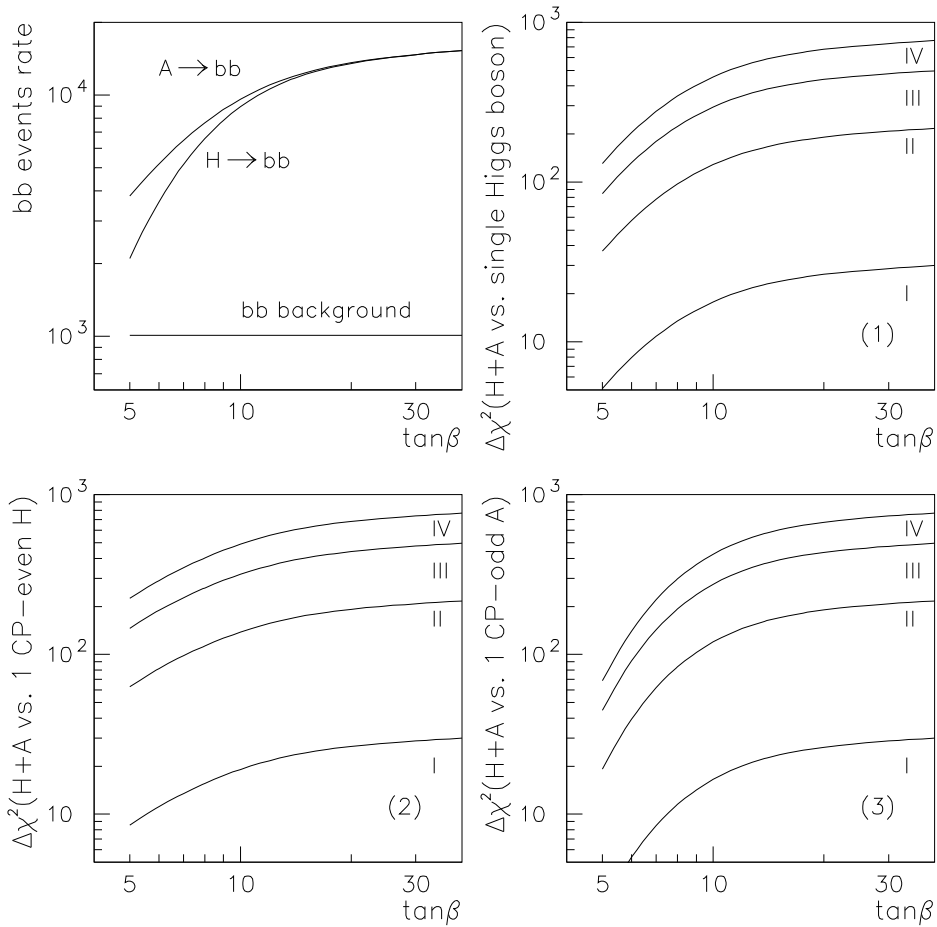}}
\medskip
\caption{
In the upper left  window, we plot the $b\anti b$ event rates
in the MSSM for the $\hh$ and $\ha$ (the $\ha$ rate is the larger
of the two) as a function of $\tanb$ for $\mha=300\gev$,
assuming squark masses of $1\tev$, no squark mixing and integrated
luminosity of $L=2\fbi$.
Also shown is the (relatively small) background rate.
In the remaining windows we plot $\Delta\chi^2$,
after including precession and increasing $L$ to $L=3\fbi$, 
as a function of $\tanb$ for three different cases
in which we forcibly lower $\mhh$ to $300\gev$ (for exact degeneracy).
(1) We adjust the $\hh$ and $\ha$ event rates so that each is exactly equal
to the average of the $\ha$ and $\hh$ rates
as predicted by the MSSM, and compute $\Delta\chi^2$ for $\hh+\ha$ vs.  
a single Higgs resonance (of any type) with the same total 
event rate, employing only the $b\anti b$ channel. 
(2) We use the actual $\hh$ event rate and compute
$\Delta\chi^2$ for $\hh+\ha$ vs. a single CP-even resonance
yielding the same $b\anti b$ event rate.
(3) As in (2), but vs. a single CP-odd resonance.
In cases (1)-(3), we give results as a function of $\tanb$
for the four polarization--luminosity
situations (i)-(iv) (as labelled on the curves) described in the text.}
\label{hh300}
\end{figure}

\begin{figure}[p]
\leavevmode
\epsfxsize=5.5in
\centerline{\epsffile{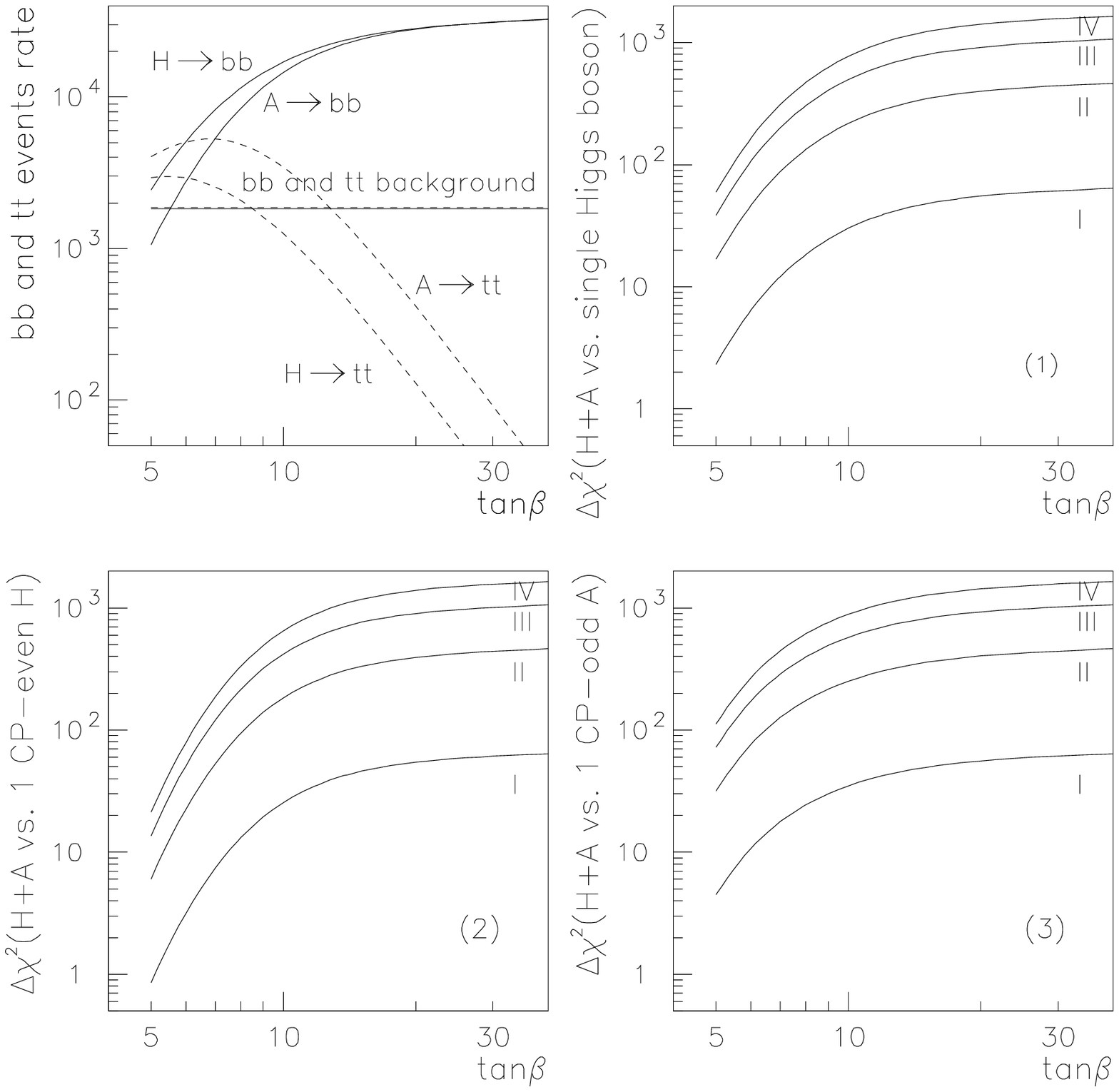}}
\medskip
\caption{The same as Fig.~\ref{hh300} except for $\mha=\mhh=400\gev$,
and using $L=7\fbi$ for the rate window increased to $L=10.5\fbi$ for the
$\Delta\chi^2$ windows (in order to account for the inefficiency
associated with spin precession).
The upper-left window gives the $\hh$ and $\ha$ and background $t\anti t$
event rates as dashed curves. In the $\Delta\chi^2$ analysis,
only the $b\anti b$ final state (after accounting for the depletion
from $\hh,\ha\to t\anti t$) is employed.}  
\label{hh400}
\end{figure}

As shown in Figure~\ref{degeneracy},
the $\hh$ and $\ha$ 
need not give exactly the same $b\anti b$ event rate (due to
differences in  $\br(\mupmum)\br(b\anti b)$). This is true
even if parameters are chosen so that they are exactly degenerate in mass
at some given $\tanb$ value.
On the other hand, it is also typically possible to choose parameters
so that they do have exactly the same rate. 
We consider this last possibility first.
(As already discussed, when the $\hh$ and $\ha$ have exactly the same
$b\anti b$ rate, there is no dependence of the summed $b\anti b$
rate on $\zeta$. As a result, 
even after including spin precession, $\Delta\chi^2$
will be independent of the CP nature of any single Higgs boson
to which we compare if we sum over the three configurations $C=I,II,III$
with luminosity weighting $L/6,L/6,2L/3$.)  
To simulate this situation, we choose a value
of $\tanb$ and a value of the $\ha$ mass. We then compute the $\hh$ and $\ha$
event rates (assuming stop masses of $1\tev$ and no squark mixing) and reset 
the $\hh$ and $\ha$ event rates so that
both are equal to the average of the originally computed $\hh$ and $\ha$ rates.
We consider $\mha$ values of $300\gev$ and $400\gev$. In the latter
case, the $t\anti t$ decay channel is open and has significant branching
ratio at lower $\tanb$ values. This, along with the decreased $\mupmum$
coupling at lower $\tanb$, 
is included in computing the $b\anti b$ rate. We make no
use of the $t\anti t$ channel; its inclusion would, of course,
increase our discrimination power, especially if the final state
correlations that can be probed there are employed.
The resulting discrimination power is generally more than adequate
without its inclusion. 

Figs.~\ref{hh300} and \ref{hh400} give
the results for $\mha=300\gev\,(400\gev)$, respectively.
We plot $\Delta\chi^2$ obtained for the four 
polarization/proton-source-intensity
options (i)-(iv) delineated earlier, but
using the higher nominal luminosities stated above: (i) $P=0.2$, 
and the $I=1$ values of $L=3\,(10.5)\fbi$;
(ii) $P=0.39$, and the $I=1$ merged-bunch $L=1.5\,(5.25)\fbi$ values;
(iii) $P=0.48$ and $I=2$, yielding merged-bunch $L=1.5\,(5.25)\fbi$;
(iv) $P=0.45$ and $I=3$, without bunch merging, yielding $L=3\,(10.5)\fbi$.
We emphasize that options (i) and (ii) require no over-design of the proton
source. The $\Delta\chi^2$ plots show that good discrimination is obtained
even for option (i) once $\tanb>10$. Option (ii)  would be
needed for good discrimination if $\tanb\sim 5$.

\begin{figure}[p]
\leavevmode
\epsfxsize=5.5in
\centerline{\epsffile{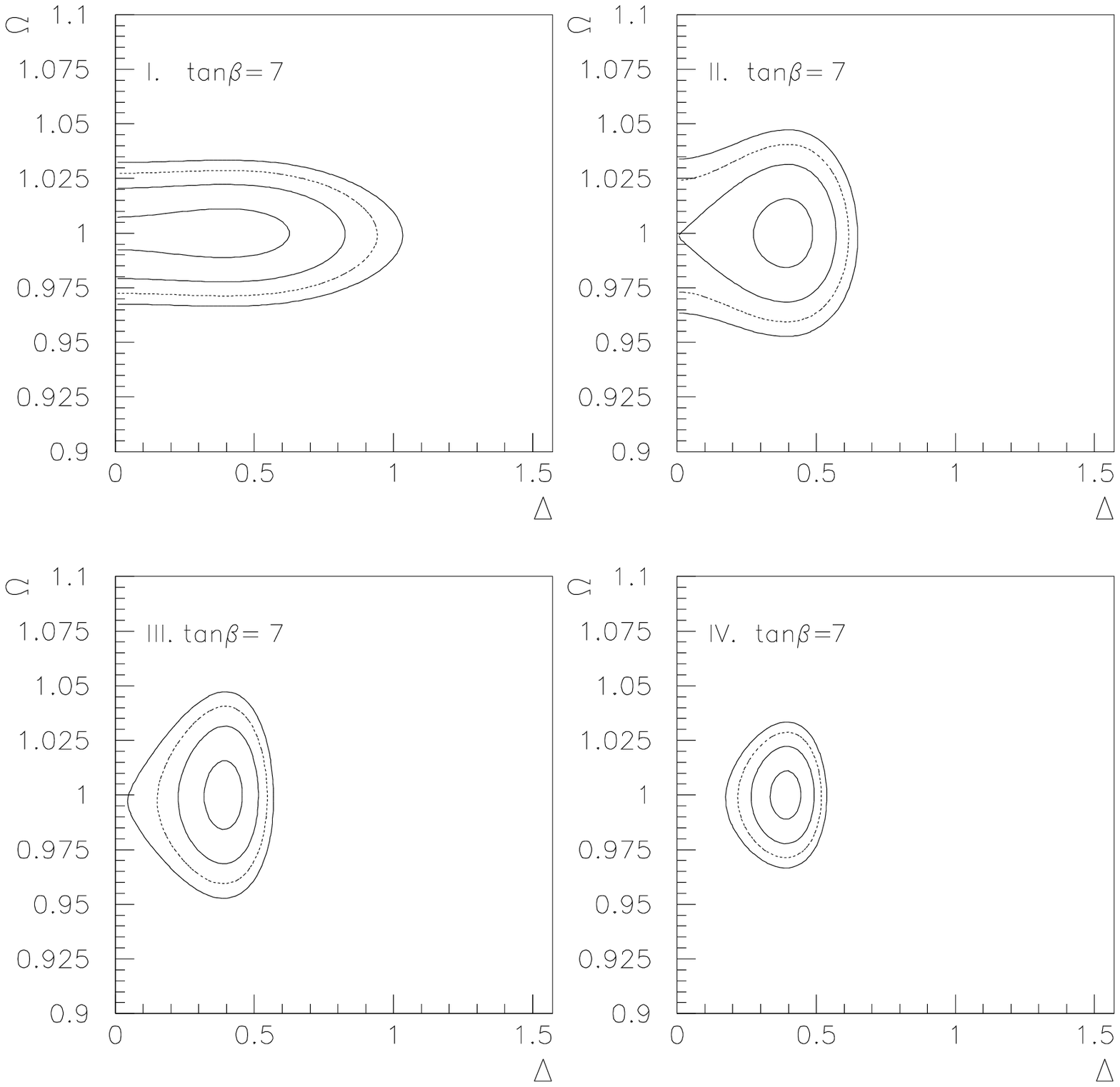}}
\medskip
\caption{We plot contours of $\Delta\chi^2=1,4,6.635,9$ in
the $\Delta,\Omega$ parameter space, assuming $\mha=400\gev$,
$\tanb=7$, squark masses of $1\tev$ and no squark mixing.
The different windows give results for the different proton source
intensity and bunch merging options described in the text.}  
\label{hh_tanb7}
\end{figure}

\begin{figure}[p]
\leavevmode
\epsfxsize=5.5in
\centerline{\epsffile{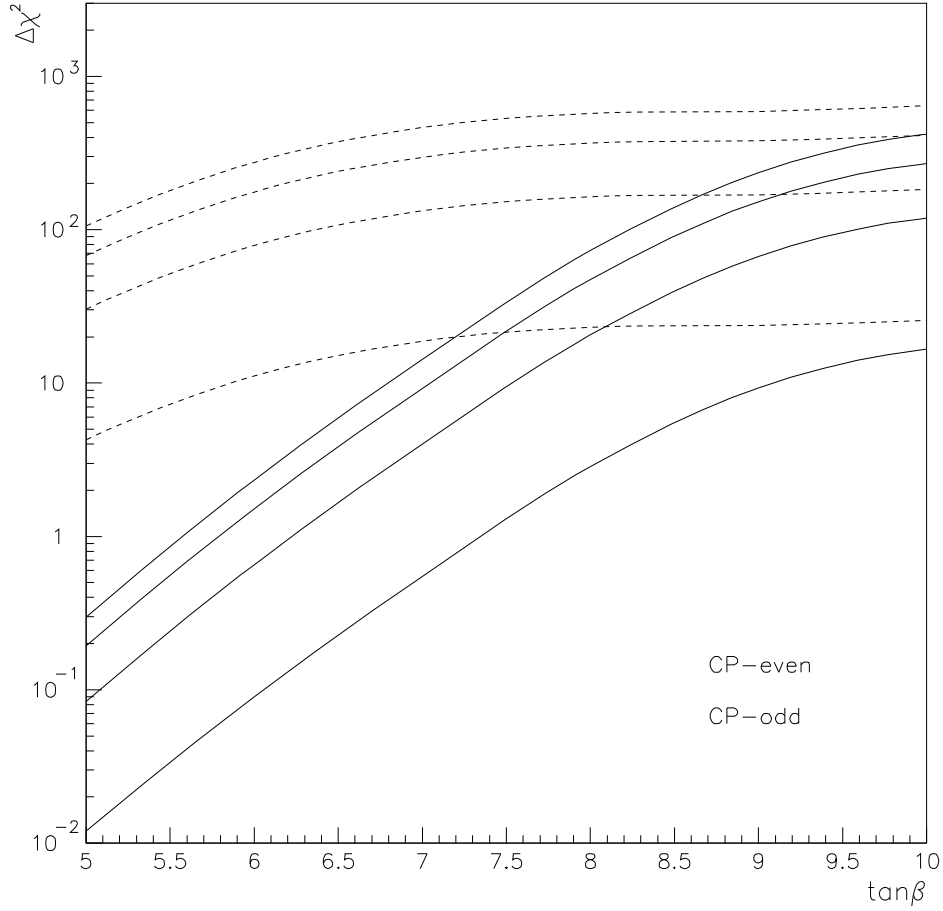}}
\medskip
\caption{We plot $\Delta\chi^2$ as a function of $\tanb$
for discriminating between the $\ha$--$\hh$ mixture predicted
by the MSSM model (as specified in the text)
at $\rts=\mhh$ compared to a single CP-even
or CP-odd Higgs boson yielding exactly the same $b\anti b$
event rate. Results are presented for the polarization/proton-source-intensity
options (i)-(iv) described in the text.}
\label{mssmchisq}
\end{figure}

We next consider the discrimination power achieved if 
we continue to enforce exact degeneracy by lowering $\mhh$
to $\mha$, but employ the 
actually predicted $\ha$ {\it and}
$\hh$ event rates.
For $\mha=300\gev$, the $\hh$ event rate for $b\anti b$
is lower than the $\ha$ rate, and the worst (best)
discrimination is achieved relative to a purely CP-odd (purely CP-even)
Higgs boson (again assumed to have exactly the same total $b\anti b$
event rate).  The resulting $\Delta\chi^2$ values for options
(i)-(iv) are given in the two bottom windows of Fig.~\ref{hh300}.
Even in the worst case, good discrimination power is achieved
for $\tanb\geq 5$ by using option (ii).
Exactly the reverse situation arises for $\mha=400\gev$. As seen in
Fig.~\ref{hh400}, the $\ha$ rate in the $b\anti b$
channel is smaller than for the $\hh$. Discrimination
against a purely CP-even Higgs will be the most difficult.
In fact, we see that option (ii) will not provide clear discrimination 
against a purely CP-even Higgs if $\tanb\lsim 7$.

Finally, let us consider the case where 
the $\hh$ and $\ha$ are not exactly degenerate,
but are only nearly so. To be specific we adopt the MSSM
predictions of Fig.~\ref{degeneracy}. As noted earlier, if the 
collider is operated with $R=0.001$ beam energy spread so as to maximize
luminosity, for $\tanb\gsim 8$ one observes a single broad peak
in a scan and we will not know that there are two Higgs bosons present.
The apparent peak of the resonance
shape will tend to coincide with the mass of the Higgs boson
which yields the higher $b\anti b$ rate (which Higgs this is depends
on the mass, as we have seen). If we center on this apparent peak,
the contribution to the $b\anti b$ rate from the weaker Higgs 
will be further reduced because $\rts$
is somewhat on the wing of its resonance shape.  This further
decreases the $\Delta\chi^2$ discrimination power relative
to a purely CP-odd (CP-even) Higgs boson
if the dominant Higgs is CP-odd (CP-even). 
Even if we see two separate peaks, we will wish to experimentally
determine which is the $\hh$ and which is the $\ha$. 

To illustrate, let us define $\Omega={\sqrt{\what N_{\ha}^2+\what N_{\hh}^2}}$
and $\Delta=\tan^{-1}{\what N_{\ha}\over \what N_{\hh}}$,
where $\what N_{\ha}$ ($\what N_{\hh}$) 
is the number of $\ha$ ($\hh$) events at the chosen $\rts$
for a possible model divided by 
$\left(N_{\ha}^2+N_{\hh}^2\right)^{1/2}_{MSSM}$.
In Fig.~\ref{hh_tanb7}, we present $\Delta\chi^2$
contours in $\Delta,\Omega$ parameter space for 
the input MSSM model (see Fig.~\ref{degeneracy})
specified by $\mha=400$, $\tanb=7$, squark
masses of $1\tev$ and no squark mixing. To determine that there are
$\ha$ events under the $\hh$ peak at 99\% CL requires the $I=2$
proton source intensity option (iii).
A global overview of our ability to discriminate the MSSM input
model from a single purely CP-even (the worst case)
or single purely CP-odd (the best case) Higgs boson yielding
exactly the same $b\anti b$ event rate is provided by Fig.~\ref{mssmchisq}.
There, we plot $\Delta\chi^2$ for these two discriminations as
a function of $\tanb$ for polarization/proton-source-intensity
options (i)-(iv).  We find that discrimination
against a purely CP-odd Higgs boson is excellent
even for $\tanb\sim 5$, whereas $\tanb>9$ [$\tanb>7.3$]
is required for 99\% CL discrimination against a single CP-even
Higgs boson using  option (i) [(ii)].
Of course, if squark mixing is large, the degree of degeneracy between
the $\hh$ and $\ha$ can be such that the peaks will merge even
when $\tanb\sim 5$. In this case, good discrimination power would
typically require enhanced proton source intensity.

\section{Summary and Conclusions}

The most natural polarization for the muon bunches at a muon
collider is $P\sim 0.2$. For this polarization, the
collider luminosity will be maximal. In initial exploration
of the Higgs or other narrow resonance, one will choose 
the polarizations to lie in the horizontal plane of the storage ring.
As the polarizations
rotate it will then be possible to perform a precise measurement
of the beam energies and their Gaussian widths and of the degree
of polarization itself. Further, by choosing
the $\mu^+$ polarization  to be 90 degrees out
of phase with the $\mu^-$ polarization, the effect of
the rotating polarization will cancel out when averaging over
the 1000 turns during which the typical bunches are stored; i.e.
the turn-averaged cross section will be identical to the
polarization-averaged cross section. However, once the basic
properties of the Higgs boson or resonance are known (total
width, branching ratios, etc.) the next important goal
will be to determine its CP properties as codified in the relative
strength of its scalar and pseudoscalar couplings to fermions.
The muon collider provides a perfect opportunity for 
determining these relative strengths for the muon itself, but only if
one has retained the ability to reconfigure the collider so
as to run with high polarizations for the muon bunches and
to have flexibility in the orientation of these polarizations
at the time the bunches are inserted into the storage ring. 
In most muon collider designs, high polarization can only be
achieved by making strong momentum cuts on the muons, which,
unless the proton source has `spare' luminosity relative
to bunch saturation limits for $P\sim 0.2$, will result in some
loss of collider luminosity. However, if the goal is to
measure the CP-odd/CP-even
coupling ratio, this loss of luminosity is more than compensated by
increased sensitivity. Our goal in this paper has been to 
develop efficient techniques and polarization configurations 
for determining the CP-odd/CP-even coupling ratio and to quantify
the accuracy with which this ratio could be extracted from the data. 

We have found that one very effective technique is to
accumulate events for three carefully chosen polarization configurations
(as defined by the polarization of the $\mu^+$ and $\mu^-$ bunches
at the time they enter the storage ring). To achieve these three
configurations, one will need appropriate solenoids and/or small rings
for manipulating the polarizations prior to injection into the storage ring.
Once the bunches are in the storage ring, further manipulation
would destroy our ability to measure the bunch energies to the 
1 part in $10^6$ level needed (at least for a narrow Higgs boson).
We have demonstrated that by selecting high-polarization muons
and performing bunch merging,
meaningful constraints on the CP nature
of the muonic couplings of a Higgs boson are possible even
if it is light and narrow and even if the proton source has
only the nominal luminosity required to saturate the bunch
limits in the storage ring at relatively low polarization.
However, we have seen that extra proton source luminosity 
(sufficient to saturate the bunch limits of the storage ring
when muon selections leading to large polarization are employed) 
may be needed to achieve a high
level of certainty regarding the CP nature of the muonic couplings
of such a Higgs boson. We have also shown that distinguishing
a highly degenerate pair of CP-even and CP-odd Higgs bosons from
a single Higgs with definite CP nature will generally be possible,
especially if this is the degenerate $\hh$--$\ha$ 
pair of the MSSM at high $\tanb$ and large $\mha$ for which event
rates are high and high-luminosity-running at $R=0.001$ would
suffice.

Overall, while the machine capabilities required 
to make a good determination of the CP nature of the muonic
couplings of a Higgs boson do not come without
a price, it could well happen that the true nature
of an observed Higgs resonance would be obscure without the
measurements considered here. Further, 
Higgs/resonance CP studies provide but one example of how unique 
sensitivity to new physics can result if we can take advantage of
the slow precession of the muon bunch polarizations by 
manipulating their orientation and relative phases and recording
events on a turn-by-turn basis. Thus, we encourage the muon
collider designers to retain the flexibilities needed to
insert devices for rotating the spins of the beams to a variety of
initial (i.e. prior to insertion into the storage ring)
configurations of longitudinal and transverse polarization
and to consider seriously the possibility of `over-designing'
the proton source relative to storage ring bunch saturation limits
associated with low polarization. Or, perhaps 
it will prove possible to employ a very high
gradient in the initial phase-rotation stage of pion/muon capture
\cite{pstareport,kaplan,geer}, thereby maximizing the luminosity available
after the selection cuts required for high polarization.

\subsection*{Acknowledgements}
We thank S Geer, R. Raja and R. Rossmanith for helpful conversations
on experimental issues.
This work was supported in part by the U.S. Department of Energy,
the U.C. Davis Institute for High Energy Physics, the State Committee for
Scientific Research (Poland) under grant No. 2~P03B~014~14 and by Maria
Sklodowska-Curie Joint Fund II 
(Poland-USA) under grant No. MEN/NSF-96-252. 
Two of the authors (BG, JP) are indebted
to the U.C. Davis Institute for High Energy Physics for the great
hospitality extended to them while this work was being performed.

\section{Appendix}

In this appendix, we present the analytic formulae for the signal
and background cross sections in an arbitrary fermionic final state.
Fermionic couplings
take the form $i\,\gamma_\mu\,e_f$ (photon), 
$i\gamma_\mu(c_f+d_f\gamma_5)$ ($Z$),
and $i(a_f+i\gamma_5 b_f)$ (Higgs). Here, $e_f=eQ_f$, 
$c_f=\frac{g}{2\cos\theta_W}(T_3^f-2Q_f\sin^2\theta_W)$, 
$d_f=-\frac{g}{2\cos\theta_W}T_3^f$, and, for a SM Higgs boson,
$a_f=\frac{gm_f}{2 m_W}$ and $b_f=0$. In terms of these quantities,
the decay width of the Higgs boson to $f\anti f$ is given by 
\begin{equation}
\Gamma(\h\to f\anti f)=\frac{1}{8\pi} \beta_f\,\mh\,(a_f^2\beta_f^2+b_f^2)\,,
\label{hwidth}
\end{equation}
where $\beta_f \equiv \sqrt{1-\frac{4 m_f^2}{s}}$.

We employ the center of mass system with $\sqrt s=E_{\rm CM}$ and define
$\zeta\equiv\zeta^+-\zeta^-$ as the angle of the $\mu^+$ transverse polarization relative to that of the $\mu^-$. 
The explicit expressions for the 4-momenta and spin vectors of the
$\mu^+$ and $\mu^-$ in the laboratory frame are:
\bea
p_{\mu^-}&=&\frac{\sqrt{s}}{2} (1,0,0, \beta)\nonumber\\
p_{\mu^+}&=&\frac{\sqrt{s}}{2} (1,0,0,-\beta)\nonumber\\
s_{\mu^-}&=& P_L^-\gamma (\beta,0,0, 1)+P_T^- (0,\cos\zeta^-,\sin\zeta^-,0)\nonumber\\
s_{\mu^+}&=& P_L^+\gamma (\beta,0,0,-1)+P_T^+ (0,\cos\zeta^+,\sin\zeta^+,0)\nonumber\\
p_f&=&\frac{\sqrt{s}}{2}
(1,  \beta_f \sin\theta \cos\phi , \beta_f \sin\theta \sin\phi,  \beta_f \cos\theta)\nonumber\\
p_{\bar{f}}&=&\frac{\sqrt{s}}{2}
(1,- \beta_f \sin\theta \cos\phi ,-\beta_f \sin\theta \sin\phi,- \beta_f \cos\theta)\,,\nonumber\\
\eea
where $\phi$ and $\theta$ are the standard angles defined 
using polar coordinates for the final fermion. The forms for
$s_{\mu^-}$ and $s_{\mu^+}$ above are those which make the separation
between longitudinal and transverse polarization most evident.
The conversion between these forms and those given earlier in Eqs.~(\ref{smum})
and (\ref{smup}) appropriate to making the precession physics most transparent
can be accomplished by the following mappings:
\bea
P_H^{\pm}\cos\theta^{\pm}=P_L^{\pm}\,,\quad
-P_H^{\pm}\sin\theta^{\pm}=P_T^{\pm}\cos\zeta^{\pm}\,,\quad
P_V^{\pm}=P_T^{\pm}\sin\zeta^{\mp}\,.
\label{mapping}
\eea

We now give expressions for the invariant matrix element squared, 
$\lvert {\cal M}\rvert^2$, for $\mupmum\to f\anti f$, summed
over final spins and averaged over initial spins.
These will be given in terms of $P_L^{\pm}$, $P_T^{\pm}$
and $\zeta^{\pm}$. Eq.~(\ref{mapping}) can be used to convert
to the $P_H^{\pm}$, $P_V^{\pm}$ and $\theta^{\pm}$ precession
variables. We divide $\lvert {\cal M}\rvert^2$ into three pieces:
the absolute square of the Higgs diagram, the interference between
the Higgs diagram and the  $\gam^*+Z^*$ exchange background diagrams,
and the absolute square of the $\gam^*+Z^*$ background diagrams.
The symbols $\Re$ and $\Im$ denote the real and imaginary parts.
We give the results for $m_\mu/\sqrt{s}\to 0$, an excellent approximation
for any reasonable Higgs mass. 
Defining $\Pi_X\equiv \left[s-m_X^2+i\Gamma_X m_X\right]^{-1}$
we have the following.

\noindent \underline{Absolute square of Higgs exchange:}
\bea
&&{\afbsqpbfsq}{\ModPihsq}\,{s^2}\,%
  \Bigl[ 
{\ambsqpbmsq} (1+P_L^- P_L^+) +  P_T^-\,P_T^+\,
  \left\{ {\ambsqmbmsq}\,{\czeta} - 2\,{\amb}\,{\bm}\,{\szeta} \right\}  \Bigr]
\nonumber\\
&&~~~~~~~~~={\afbsqpbfsq}\,{\ModPihsq}\,{s^2}\,{\ambsqpbmsq}%
  \left[ 
 (1+P_L^- P_L^+) +  P_T^-\,P_T^+\,\cos(\zeta+2\delta)\right]
\label{higgssq}
\eea

\noindent\underline{Higgs -- $\gam^*+Z^*$ Interference:}
\newcommand{\PigPihs}{(\Pig \Pi^*_h)}
\newcommand{\PizPihs}{(\Pi_Z \Pi^*_h)}
\bea
  4 \sin\theta \,a_f {\ambsqpbmsq}^{1/2} m_f \beta_f s^{3/2} \Bigl\{ 
\hspace{8cm}
\nonumber\\
    -\sin(\zeta^++\delta-\phi)\,P_T^+\,\left[ 
      e_f\,e_\mu\,\Im\PigPihs+c_f\,(c_\mu+d_\mu\,P_L^-)\,\Im\PizPihs 
    \right]
\nonumber\\
    -\sin(\zeta^--\delta-\phi)P_T^- \,\left[ 
      e_f\,e_\mu\,\Im\PigPihs+c_f\,(c_\mu-d_\mu\,P_L^+)\,\Im\PizPihs 
    \right]
\nonumber\\ 
    -\cos(\zeta^++\delta-\phi)\,P_T^+\,\left[ 
      e_f\,e_\mu\,P_L^-\,\Re\PigPihs 
      +c_f\,(c_\mu\,P_L^- + d_\mu)\,\Re\PizPihs 
    \right]\phantom{\Bigr\}}
\nonumber\\
    +\cos(\zeta^--\delta-\phi)\,P_T^-\,\left[ 
      e_f\,e_\mu\,P_L^+\,\Re\PigPihs 
      + c_f\,(c_\mu\,P_L^+ - d_\mu)\,\Re\PizPihs 
    \right]
  \Bigr\}\nonumber\\
\eea

\noindent\underline{Absolute square of $\gam^*+Z^*$:}
\bea
&&-8\,d_f^2\Bigl[ 
  \left( {\cmusq} + {\dmsq} \right)\,\left(1-P_L^- P_L^+\right)
+2\,c_\mu\,d_\mu\,\left(P_L^-- P_L^+\right) 
\Bigr]\,m_f^2\,\ModPizsq\,s\,\nonumber\\
&&+2\,\sin(\zeta-2\phi)\,\sin^2\theta\,P_T^-\,P_T^+{c_f}\,
     d_\mu\,{e_f}\,{e_\mu}\,{\ImPiz}\,\Pig\,\beta_f^2\,s^2\nonumber\\ 
&&+\cos(\zeta-2\phi)\,\sin^2\theta\,P_T^- P_T^+\,\Bigl\{ 
  \left( {\cfsq} + {\dfsq} \right)\,\left( {\cmusq} -{\dmsq} \right)\,\ModPizsq
+ {e_f^2}\,{e_\mu^2}\,\Pig^2 +
  2 {e_f}\,{e_\mu}\,{c_f}\,c_\mu\,\Pig\,\Re(\Pi_Z) 
\Bigr\}\,\beta_f^2\,s^2\nonumber\\ 
&&+\left(2-\beta_f^2 \sin^2\theta\right)\,\Bigl\{
  \left( {\cfsq} + {\dfsq} \right)\,\left[
    \left( {\cmusq} + {\dmsq} \right)\,
    \left(1-P_L^- P_L^+\right)+2\,c_\mu\,d_\mu\,
    \left(P_L^-- P_L^+\right)
  \right]\,\ModPizsq
 \nonumber\\
&&
  +2\,{c_f}\,{e_f}\,{e_\mu}\,
  \left[ 
    c_\mu\,\left(1-P_L^- P_L^+\right)+d_\mu\,
    \left(P_L^-- P_L^+\right)
  \right]\,\Pig\,\Re(\Pi_Z)
  +{e_f^2}\,{e_\mu^2}\,\left(1-P_L^- P_L^+\right)\,\Pig^2
\Bigr\} \,s^2
\nonumber\\
&&+4\,\cos\theta\,d_f
\Bigl\{
   c_f\,\left[ 
      2\,c_\mu\,d_\mu\,\left(1-P_L^- P_L^+\right) + 
      \left( {\cmusq} + {\dmsq} \right)\,\left(P_L^-- P_L^+\right)
   \right]\,\ModPizsq  
 \nonumber\\ 
&&
  +e_f\,e_\mu\,\left[
    d_\mu\,\left(1-P_L^- P_L^+\right)+c_\mu\,\left(P_L^-- P_L^+\right) 
  \right]\,\Pig\,\Re(\Pi_Z)
\Bigr\}\,\beta_f\,s^2
\eea

In terms of $|{\cal M}|^2$, the cross section as a function of $s$, $\theta$
and $\phi$ is given by
\beq
{d\sigma\over d\cos\theta\, d\phi}={1\over 64\pi^2 s}\beta_f |{\cal M}|^2\,.
\label{xsec}
\eeq 
For example, if we spin average the Higgs portion of $|{\cal M}|^2$
by combining $\zeta=0$ and $\zeta=\pi$ (so that the average of $\cos\zeta$
is zero),
integrate over $d\cos\theta$ and $d\phi$, and use Eq.~(\ref{hwidth}),
then we obtain the standard spin-averaged total cross section:
$\sigma(s)=4\pi\Gamma(\h\to\mupmum)\Gamma(\h\to f\anti f)\lvert\Pi_h\rvert^2$.
When this latter form is convoluted with a Gaussian distribution
in $\rts$, one obtains the result of Eq.~(\ref{sig0form}).

\end{document}